\documentclass{aa} 
\usepackage{amsmath} 
\usepackage{txfonts}
\usepackage{graphicx} 
\usepackage{url}
\usepackage{natbib} 
\bibpunct{(}{)}{;}{a}{}{,}

\newcommand{\ie}{{\it i.e.}~}

\newcommand{\Ms}{M$_\odot$}

\newcommand{\el}[2]{$\rm{}^{#2}\kern-0.6pt#1$}
\newcommand{\del}[1]{\delta\text{#1}}
\usepackage{color}

\newcommand{\modif}[1]{#1}

\begin{document}

\title{Fast rotating massive stars and the origin of the 
abundance patterns in galactic globular clusters}


\author{T. Decressin\inst{1} \and G. Meynet\inst{1} \and
  C. Charbonnel\inst{1,2} \and N. Prantzos\inst{3} \and
  S. Ekstr\"om\inst{1}}

\offprints{T. Decressin,\\ \email{Thibaut.Decressin@obs.unige.ch}}

\institute{Geneva Observatory, University of Geneva,
  chemin des Maillettes 51, 1290 Sauverny, Switzerland
\and 
  Laboratoire d'Astrophysique de Toulouse et Tarbes,
  CNRS UMR 5572, OMP, 14, Av. E.Belin, 31400 Toulouse, France
\and
  Institut d'Astrophysique de Paris, CNRS UMR 7095, Univ. P. \& M.Curie, 
98bis Bd. Arago, 75104 Paris, France \\
}

\date{Received / Accepted}

\authorrunning{} \titlerunning{}

\abstract{} %
{We propose the Wind of Fast Rotating Massive Stars scenario to explain the
  origin of the abundance anomalies observed in globular clusters.}  {We
  compute and present models of fast rotating stars with initial masses
  between 20 and 120~\Ms{} for an initial metallicity $Z$=0.0005
  ($\text{[Fe/H]}\simeq-1.5$).  We discuss the nucleosynthesis in the H-burning core of
  these objects and present the chemical composition of their ejecta. We
  consider the impact of uncertainties in the relevant nuclear reaction
  rates.}  {Fast rotating stars reach the critical velocity at the
  beginning of their evolution and remain near the critical limit during
  the rest of the main sequence and part of the He-burning phase. As a
  consequence they lose large amounts of material through a mechanical wind
  which probably leads to the formation of a slow outflowing disk. The
  material in this slow wind is enriched in H-burning products and presents 
  abundance patterns similar to the chemical anomalies observed in globular
  cluster stars. In particular, the C, N, O, Na and Li variations are
  well reproduced by our model. However the rate of the
    \el{Mg}{24}$(p,\gamma)$ has to be increased by a factor 1000 around
    $50\times 10^6$~K in order to reproduce the whole
    amplitude of the observed Mg-Al anticorrelation.
  We discuss how the long-lived low-mass stars
  currently observed in globular clusters could have formed out of the slow
  wind material ejected by massive stars.} {}


\maketitle

\section{Introduction}

Galactic globular clusters (hereafter GCs) appear to be chemically
homogeneous (with the notable exception of $\omega$ Cen) with respect to
the iron-group (Mn, Fe, Ni, Cu), neutron-capture (Ba, La, Eu) and
alpha-elements (Si, Ca) (e.g.,
\citealt{KraftSneden1992,JamesBonifacio2004a,JamesBonifacio2004b,Sneden2005,SobeckIvans2006}).
However it has long been known that these large aggregates of stars show
strong inhomogeneities in lighter elements: C, N, O, Na, Mg and Al
abundances present indeed large star-to-star abundance variations within
all the individual GCs studied up to now (for complete references see the
early reviews by \citealt{FreemanNorris1981}, \citealt{Smith1987}, and
\citealt{Kraft1994} and the more recent ones by \citealt{GrattonSneden2004}
and \citealt{Charbonnel2005}).

The observed patterns point to the simultaneous operation of the CNO, NeNa
and MgAl cycles of hydrogen-burning : C and N, O and Na, and Mg and Al are
respectively anticorrelated, the abundances of C, O and Mg being depleted
while those of N, Na and Al are enhanced. Whenever C, N, and O are observed
simultaneously, their sum appears to be constant within the observational
errors \citep[e.g.,][]{DickensCroke1991,IvansSneden1999}. The sum Mg+Al is
also found to be constant in several clusters \citep{Shetrone1996II}.
Observations in NGC~6752, M~13 and M~71 show that the Mg depletion is due
to the burning of \el{Mg}{24} while \el{Mg}{25} is untouched and
\el{Mg}{26} is produced in the Al-rich stars
\citep{YongGrundahl2003,YongGrundahl2005,YongAoki2006}. Last but not
  least, Li was found to be anticorrelated with Na in turn-off stars of
  NGC~6752 \citep{PasquiniBonifacio2005}. All these features are
considered anomalous because they are not seen in field stars of similar
metallicity \citep[e.g.,][]{GrattonSneden2000}.

For many years only the brightest GC red giants were accessible for
detailed spectroscopic observations, and two main theoretical streams were
competing to explain the available data : (1) the so-called ``evolution''
scenario according to which the chemical anomalies are generated inside the
low-mass stars we are presently observing, and (2) the ``self-enrichment''
(or primordial) scenario according to which such patterns pre-existed in
the protocluster gas and were inherited at the birth of the long-lived
stars.

The evolution hypothesis has been seriously challenged by recent
spectroscopic observations of less luminous stars in earlier stages of
evolution in a number of GCs. Such studies revealed indeed that stars
located slightly above and below the main sequence turnoff exhibit the same
anomalies as their giant counterparts
\citep{GrattonBonifacio2001,GrundahlBriley2002,CarrettaBragaglia2003,CarrettaBragaglia2004,CohenBriley2002,RamirezCohen2002,RamirezCohen2003,HarbeckSmith2003}.
However such objects are not hot enough\footnote{In the central region of a
  0.85~\Ms, [Fe/H] = -1.3 turnoff star, the temperature is of the order of
  $25 \times 10^6$~K.} for the required set of nuclear reactions to occur
within their interior. Let us recall that while the CNO cycle is activated
for temperatures above $20 \times 10^6$~K, the NeNa and MgAl chains require
temperatures around $35 \times 10^6$~K and $50 \times 10^6$~K respectively.
Destruction of \el{Mg}{24} by proton-capture needs still higher
temperatures, around $70 \times 10^6$~K (e.g.,
\citealt{ArnouldGoriely1999}; Prantzos \& Charbonnel in preparation). As a
consequence the abundance variations cannot be produced in situ, but
certainly reflect the initial composition of the protostars. It is thus
clear now that a large fraction of GC low-mass stars were formed from
material processed through H-burning at high temperatures and then lost by
more massive and faster evolving stars, and perhaps mixed with some
original gas. Various aspects of this ``self-enrichment scenario'' are
discussed in details by \citet[ hereafter PC06]{PrantzosCharbonnel2006}.

Regarding the nucleosynthetic site, most studies have focused on massive
AGB stars which were suggested as the possible polluters by
\citet{CottrellDaCosta1981}. The two main reasons why these objects have
been favored are that (1) they host regions where H-burning occurs at high
temperatures (in particular when they experience the so-called hot bottom
burning, or HBB, at the base of their convective envelope between 
successive thermal pulses), and (2) the material they eventually eject (by
stellar winds or Roche lobe overflow) is not enriched in iron. This last
property is well consistent with the observational fact recalled above that
the iron abundance in a GC does not show any significant scatter. For long
the AGB hypothesis was discussed only on a qualitative basis. However
recent custom-made stellar models
(\citealt{VenturaDAntona2001,VenturaDAntona2002,DenissenkovHerwig2003,KarakasLattanzio2003,Herwig2004a,Herwig2004b,VenturaDAntona2005a,VenturaDAntona2005b,VenturaDAntona2005c,VenturaDAntona2006};
Decressin et al., in preparation) pointed out very severe difficulties from
the nucleosynthesis point of view which stem from the competition between the
HBB and the third dredge-up. This latest process does indeed contaminate
the envelope of the AGB with the products of helium burning and creates
abundance patterns in conflict with the observed ones (see
\citealt{FennerCampbell2004} and \citealt{Charbonnel2005} for more
details).  PC06 discuss other shortcomings of the AGB scenario, the main
one being related to the peculiar initial mass function it requires. In
addition, they underline the fact that the AGB scenario gives no
satisfactory answer as to the role of stars more massive and less massive
than the presumed polluters.

Massive stars, more precisely, Wolf-Rayet stars, have been proposed
  by \citet{BrownWallerstein1993} and by \citet{WallersteinMyckkyLeep1987}
  as possible sources for the very early enrichment of globular clusters.
  More recently \citet{MaederMeynet2006} suggested that He-rich stars in
  $\omega$ Cen could be formed from wind material of fast rotating massive
  stars. \citet{PrantzosCharbonnel2006} proposed a comprehensive (albeit
  qualitative) scenario for the role of massive stars, suggesting that
  their winds provide the metal-enriched material for the next stellar
  generation, and that the subsequent supernova explosions provide the
  trigger for the star formation; the SN ejecta escape altogether the GC
  environment, along the cavities opened previously by the stellat winds.
  PC06 also studied the massive star IMF required to explain quantitatively
  the amount of Na-enhanced stars observed in NGC~2708 and found it to be
  flatter than canonical (i.e. Salpeter) IMFs; even flatter IMFs would be
  required in case the polluters were AGB stars. Similar conclusions for a
  flat IMF are reached by \citet{Smith2006}, for the case of N enhancement
  of GC by massive star winds.
The main reason why such objects have been discarded in
the past in the context of the self-enrichment scenario is related to the
fact that iron is ejected at the time of their supernova explosion and this
constitutes a priori a serious drawback for considering massive stars as GC
pollution sources. This is true unless some filtering process
removes the ejecta enriched by helium burning and more advanced nuclear
stages while preserving those bearing the signatures of hydrogen
processing. This is the key point of the {\sl Winds of Fast Rotating Massive
  Stars} scenario (hereafter WFRMS) that we propose in the present work. In
our framework, the GC chemical anomalies are built in H-burning zones of
massive stars.  Rotational mixing brings to the surface CNO-processed
material which can then be ejected in a slow wind when the stars rotate at
the critical limit.

As shown by \citet{SackmannAnand1970} and later by \citet{Langer1998} and
\citet{MaederMeynet2001}, massive stars do reach the so-called critical
velocity\footnote{By critical velocity, we mean the equatorial surface
  velocity such that the centrifugal acceleration exactly balances the
  gravity.} early on the main sequence if (1) they start their evolution
with a sufficiently high initial rotation rate, (2) they do not lose too
much angular momentum through stellar winds and (3) an efficient mechanism
(meridional circulation in our models) transports angular momentum
from the core to the envelope. Once the critical limit is reached,
the surface velocity remains near the critical value during the rest of the
main sequence and very likely an equatorial disk is formed as
observed for instance around Be stars. In the case of these stars, only
minor outflow in the line-forming region is observed (see the review by
\citealt{PorterRivinius2003}). This material has thus a great chance to be
retained in the GC potential well. On the other hand fast rotation leads to
strong internal mixing of the chemicals. As a result the ejected material
will present the marks of H-processing occurring in the stellar core. Thus
if new stars form out of the slow wind material, their composition would bear 
the signatures of H-burning. The main question addressed in this paper is 
whether the chemical composition of the ejecta of fast rotating massive stars 
is compatible with what observed in the long-lived GC stars. 

In the present paper we develop in details the WFRMS scenario. In \S~2 we
describe the physical ingredients of our models of rotating massive stars.
We first focus on the properties of the 60~\Ms{} models computed with
various assumptions. For this initial stellar mass we discuss the
nucleosynthesis in the H-burning core in \S~3 while in \S~4 we investigate 
the chemical composition of the wind ejecta. The full range of masses between
20 and 120~\Ms{} is investigated in \S~5. In \S~6 comparisons
between the wind composition and the GC abundance patterns are performed. A
schematic and speculative discussion of the complete scenario for
explaining the inhomogeneities in GCs is presented in \S~7. 
The conclusions and some future lines of research are given in \S~8.

\section{Physical inputs}

\begin{figure}[htbp]
  \includegraphics[width=0.4\textwidth]{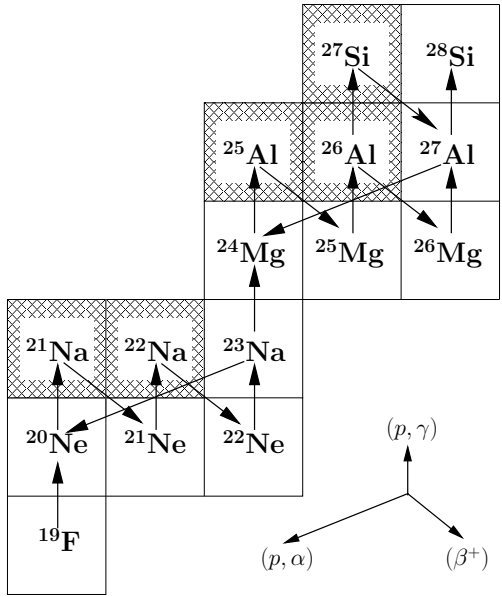}
  \caption{Network of nuclear reactions involved in the NeNa and MgAl chains. 
    Unstable nuclei are in shaded boxes. Arrows indicate the type of
    nuclear reactions: $(p,\gamma)$, $\beta^+$, and $(p,\alpha)$}
  \label{fig:net}
\end{figure}

Our stellar models were computed with the Geneva evolution code including
the effects of rotation \citep{MeynetMaeder2000}. We focus on the mass range
corresponding to stars with high enough central temperatures on the main
sequence for the NeNa and MgAl chains (see Fig.~\ref{fig:net}) to be
activated. Models with initial masses of 20, 40, 60, and 120~\Ms{} are computed
from the Zero Age Main Sequence up to the end of the core He-burning
phase, and a 200~\Ms{} model is computed up to the end of the core
H-burning phase. The theoretical predictions for the 60~\Ms{} star are
discussed in detail in \S~3 and 4 in order to properly illustrate the WFRMS
scenario.

\subsection{Microphysics}

We use the OPAL opacities \citep{IglesiasRogers1996}, complemented at
temperatures below 5000~K with the molecular opacities of
\citet{AlexanderFerguson1994}
(\url{http://webs.wichita.edu/physics/opacity}).

\begin{table}[htbp]
  \caption{Main physical inputs of the stellar models for the various
    initial masses considered. The labels ``r'' and ``s'' indicate respectively the
    models computed with or without rotation while the labels A-D refer to
    the choices in nuclear reactions (see text and Table~\ref{tab:NR}). The
    initial value of $\Omega/\Omega_\text{crit}$ is given. [I], [H02] and
    [H04] correspond respectively to \citet{IliadisStarrfield2001} and
    \citet{HaleChampagne2002,HaleChampagne2004}}  
  \label{tab:PI}
  \begin{tabular}{cccl}
    \hline
    M (\Ms) & Label & $\Omega/\Omega_\text{crit}$ & Nuclear rates \\
    \hline
    \hline
    60 & 60rA & 0.95 & set A$^\dag$: NACRE (nominal)   \\
    & 60rB & 0.95 & set B$^\dag$: [I, H02, H04] (nominal) \\
    & 60rC & 0.95 & set C$^\dag$: [I, H02, H04] (exp. limits) \\
    & 60rD & 0.95 & set D$^\dag$ \\
    & 60rE & 0.80 & set C \\
    \hline
    \hline
    20 & 20rC & 0.95 & set C \\
    40 & 40rC & 0.98 & set C \\
    120 & 120rC & 0.80 & set C \\
    200 & 200rC & 0.95 & set C \\
    20 & 20sC & 0 & set C \\
    40 & 40sC & 0 & set C \\
    60 & 40sC & 0 & set C \\
    120 & 120sC & 0 & set C \\
    \hline
  \end{tabular}
  
  $^\dag$ Details on the nuclear rates used are presented in
  Table~\ref{tab:NR}.

\end{table}

\begin{table}[htbp]
  \caption{Nuclear reaction rates adopted for the NeNa- and MgAl-chains in
  the sets B-D. [N], [I], [H02] and [H04] correspond respectively to
  \citet{AnguloArnould1999}, \citet{IliadisStarrfield2001} and
  \citet{HaleChampagne2002,HaleChampagne2004}. ``Nom.'',
  ``low.'' and ``up.'' refer respectively to nominal, lower and upper
  limits of the experimental values. For all the other reactions of our
  network we use the NACRE nominal values }
  \label{tab:NR}
  \begin{tabular}{ccccc}
    \hline
    Reaction & set A & set B & set C & set D \\
    \hline
    \hline
    \el{Ne}{20}($p,\gamma$) & [N], nom. & [N], nom. & [N], low.  & [N], low. \\
    \el{Ne}{21}($p,\gamma$) & [N], nom. & [I], nom. & [I], low.  & [I], low. \\
    \el{Ne}{22}($p,\gamma$) & [N], nom. & [H2], nom. & [H2], low. & [H2], low. \\
    \el{Na}{23}($p,\gamma$) & [N], nom. & [H4], nom. & [H4], low. & [H4], low. \\
    \el{Na}{23}($p,\alpha$) & [N], nom. & [H4], nom. & [H4], up. & [H4], up. \\
    \el{Mg}{24}($p,\gamma$) & [N], nom. & [I], nom. & [I], up.   & [I], +3~dex  \\
    \el{Mg}{25}($p,\gamma$) & [N], nom. & [I], nom. & [I], up.   & [I], up. \\
    \el{Mg}{26}($p,\gamma$) & [N], nom. & [I], nom. & [I], up.   & [I], up. \\
    \el{Al}{27}($p,\gamma$) & [N], nom. & [I], nom. & [I], low.  & [I], low. \\
    \el{Al}{27}($p,\alpha$) & [N], nom. & [I], nom. & [I], low.  & [I], low. \\
    \hline
  \end{tabular}
\end{table}

\begin{figure*}[htbp]
  \includegraphics[width=0.50\textwidth]{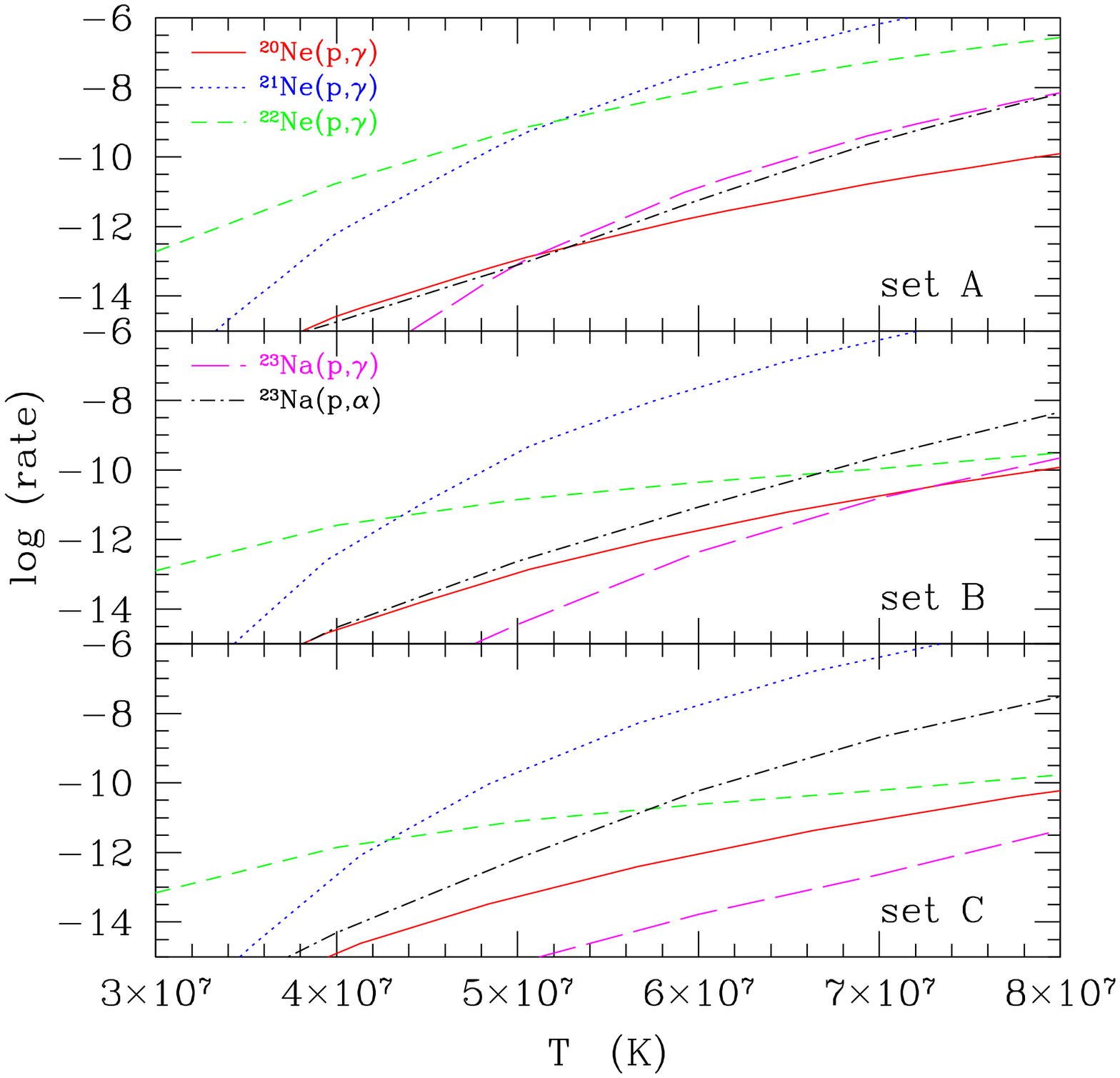}
  \includegraphics[width=0.50\textwidth]{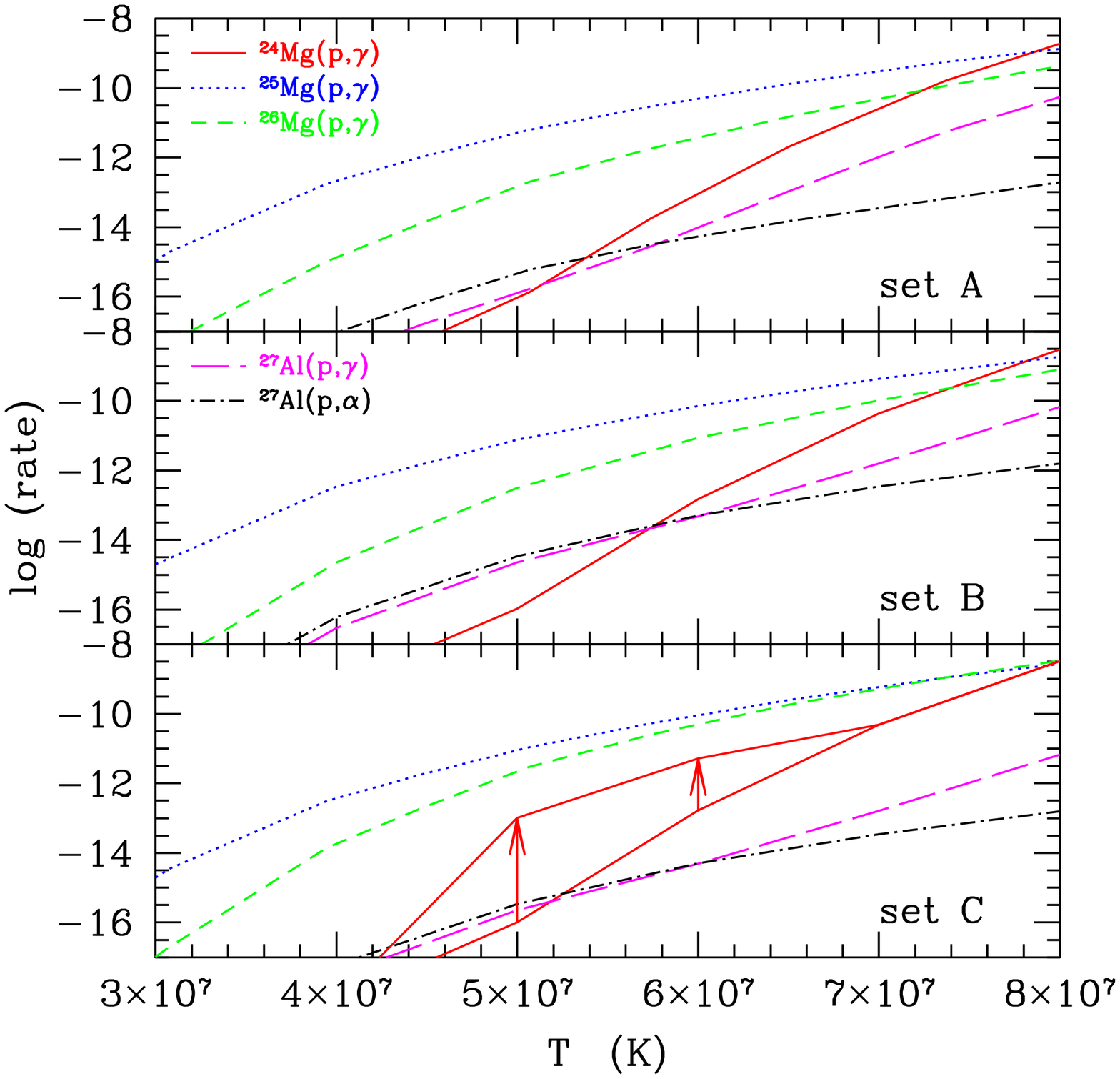}
  \caption{Nuclear reaction rates for NeNa (left) and MgAl chains (right)
    from the set A, B and C (top to bottom). Arrows and upper full 
    line in the lower right panel indicate the increase of the rate of
    \el{Mg}{24}(p,$\gamma$)$^{25}$Al at $T=50$ and $60\times 10^6$~K
      assumed in set D.}
  \label{fig:NR}
\end{figure*}

In order to investigate the effects of the nuclear reaction rate
uncertainties on the products of hydrogen nucleosynthesis, we present four
models for the 60~\Ms{} star computed using different sets of nuclear
reactions for the hydrogen-burning network (see Table~\ref{tab:PI}). Set~A
uses all the nominal values of the NACRE compilation
\citep{AnguloArnould1999}. The three other cases include the updates of
\citet{IliadisStarrfield2001} and \citet{HaleChampagne2002,
  HaleChampagne2004} for the reactions involved in the NeNa and MgAl chains
but with different options (see Table~\ref{tab:NR}). Set~B includes nominal
values while in set~C some specific rates are set to the experimental upper
or lower limits.  Fig.~\ref{fig:NR} presents the corresponding rates for
the temperature range between $30 \times 10^6$ and $80 \times 10^6$~K which
is typical of the central temperatures of our main sequence stars.  Finally
set~D is similar to set~C except for the proton-capture on \el{Mg}{24}
which is increased by three orders of magnitude compared to
\citet{IliadisStarrfield2001} nominal value around $50\times 10^6$~K. 

The initial composition of the chemical mixture is given in
Table~\ref{tab:Ab}. It corresponds to that used to compute the opacity
tables (\citealt{IglesiasRogers1996}, Weiss alpha-enhanced elements
mixture). The metallicity of our models is $\text{[Fe/H]} \simeq -1.5$
corresponding to that of NGC~6752 which is the GC with the largest set of
abundance data. The initial isotopic ratios of magnesium are taken equal to
80:10:10; this corresponds to the values observed in NGC~6752
``unpolluted'' stars (i.e., in stars with high O and low Na abundances) in
contrast with ``polluted'' stars which display large O depletion with high
Na abundance \citep{YongGrundahl2003,YongAoki2006}.

\begin{table}[htbp]
  \caption{Initial abundances in mass fraction}
  \label{tab:Ab}
  \begin{tabular}{cccc}
    \hline
    Element & Abundance & Element & Abundance \\
    \hline
    \hline
    \el{H}{1}   & 0.754     & \el{F}{19}  & 1.53e-8 \\
    \el{He}{3}  & 2.93e-5   & \el{Ne}{20} & 5.30e-5 \\
    \el{He}{4}  & 0.245     & \el{Ne}{21} & 5.00e-8 \\
    \el{C}{12}  & 3.50e-5   & \el{Ne}{22} & 4.72e-6 \\
    \el{C}{13}  & 1.47e-7   & \el{Na}{23} & 3.30e-7 \\
    \el{N}{14}  & 1.03e-5   & \el{Mg}{24} & 1.68e-5 \\
    \el{N}{15}  & 1.58e-8   & \el{Mg}{25} & 2.10e-6 \\
    \el{O}{16}  & 3.00e-4   & \el{Mg}{26} & 2.10e-6 \\
    \el{O}{17}  & 1.31e-7   & \el{Al}{27} & 9.00e-7 \\
    \el{O}{18}  & 7.45e-6   & \el{Si}{28} & 2.56e-5 \\
    \hline
  \end{tabular}
\end{table}

\subsection{Rotation and mass loss}

We follow the formalism by \citet{Zahn1992} and \citet{MaederZahn1998} for
the transport of angular momentum and chemicals in rotating stars.  The
effects of both meridional circulation and shear turbulence are taken into
account: the meridional circulation advects angular momentum and the shear
acts as a diffusive process. The transport of chemical species is computed
as a diffusive process as the result of meridional circulation and
horizontal and vertical turbulence \citep{ChaboyerZahn1992}.  The treatment
of the convective instability is done according to the Schwarzschild
criterion and we do not consider overshooting.

The treatment of rotation includes the hydrostatic effects following
\citet{MeynetMaeder1997} as well as the impact of rotation on the mass loss
rate described by \citet{MaederMeynet2000}. We do not account for the wind
anisotropies induced by rotation as in \citet{Maeder1999} although the
related effects would reinforce the trends found in this paper by fastening
the arrival at the break-up limit (see \S~4.1).

The radiative mass loss rates are from \citet{KudritzkiPuls2000} when $\log
T_{\rm eff} > 3.95$ and from \citet{deJagerNieuwenhuijzen1988} otherwise.
When a model reaches the WR phase (i.e., when the surface hydrogen mass
fraction becomes lower than 0.4 and the effective temperature is higher
than $10^4$~K), the mass loss rate is switched to the prescription of
\citet{NugisLamers2000}. Except for the WR phase, we consider a dependence
of the mass loss rates with metallicity as $\dot{M} \propto
\sqrt{Z/Z_{\odot}}$, where $Z$ is the mass fraction of heavy elements at
the surface of the star.

As in \citet{MeynetMaeder2006} a specific treatment for mass loss has been
applied at break-up. 
Let us recall that according to \citet{MeynetMaeder2000} 
three kinds of ``break--up limits'' can be defined depending on which
mechanism (i.e., radiative acceleration, centrifugal acceleration
or both) contributes to counterbalance the gravity: 1.-- The
$\Gamma$--Limit, when radiation effects largely dominate;
2.-- The $\Omega$--Limit, when
rotation effects are essentially determining break--up;
3.-- The $\Omega \Gamma$--Limit, when both
rotation and radiation are important for the critical velocity.
In the present work the $\Omega$--Limit is reached during the MS,
while the $\Omega \Gamma$--Limit is encountered by our most massive stellar models
($M \ge 60$ M$_\odot$) just after the MS.

During the MS, near the $\Omega$--Limit, two counteracting effects are
competing.  On one hand, matter is removed from the stellar surface
(mainly through the equatorial regions) together with angular momentum.
Also the expansion of the envelope tends to slow down the surface. On the
other hand, meridional advection in the outer layers acts so as to transfer
angular momentum from the inner stellar regions to the surface (see for
instance Fig.~1 in \citealt{MeynetMaeder2002}). This acts as to accelerate
the surface. As long as the internal transport of angular momentum is
efficient enough for accelerating the outer layers in a timescale shorter
than the mass loss or inflation timescale, the surface velocity remains
near the critical limit.  In practice, however, the critical limit contains
mathematical singularities; we thus consider that during the break-up
phase, the mass loss rate is such that the rotation velocity stays near a
constant fraction of the critical value (0.98 typically). At the end of the
main sequence, the stellar radius inflates so rapidly that meridional
circulation cannot anymore ensure the internal coupling and the break-up
phase ceases naturally.  However the most massive stellar models
  encounter the $\Omega \Gamma$--Limit after the MS. In that case, we apply
  similar procedures as the one described above, \ie we apply mass loss
  rates that maintain the model at a constant distance from the $\Omega
  \Gamma$--Limit. When the star has lost sufficient amount of mass for
  evolving away from this limit, the regular mass loss rates apply again.

For the reasons invoked in the introduction and which will be discussed
again later, we explore the case of high initial rotational velocities.
For the stars with initial mass between 40 and 120~\Ms{} we take
$V_\text{ini} = 800$~km~s$^{-1}$. The 200~\Ms{} model starts with
$V_\text{ini} = 1000$~km~s$^{-1}$. This corresponds to an initial value of
$\Omega/\Omega_{\rm crit}$ between $0.80$ and $0.98$. The 20~\Ms{} star
starts with $V_\text{ini} = 600$~km~s$^{-1}$ as it nearly corresponds to
the break-up velocity ($\Omega/\Omega_{\rm crit}= 0.95$).

\section{Central hydrogen-burning in a 60~\Ms{} star}

\begin{figure}[htbp]
  \includegraphics[width=0.48\textwidth]{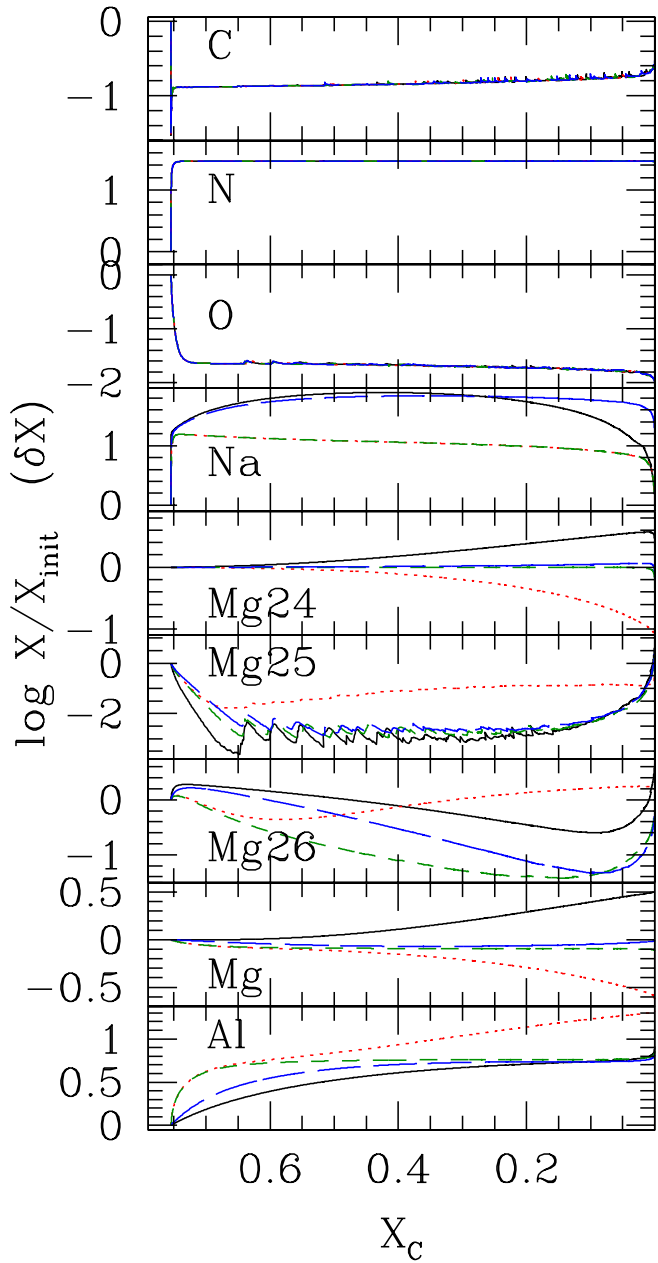}
  \caption{Evolution of the central abundances in the 60~\Ms{} 
    rotating models computed with different sets of nuclear reaction rates:
    set~A (full line), set~B (long dashed lines), set~C (short dashed
    lines) and set~D (dotted lines). The amount of the unstable nucleus
    \el{Al}{26} is added to those of \el{Mg}{26} and Mg.}
  \label{fig:AbC}
\end{figure}

As a prerequisite to our scenario we first have to check whether the
abundance patterns due to nuclear reactions in the hydrogen-burning core of
massive stars do mimic the chemical trends observed in GC low-mass stars.
Obviously the abundance variations obtained within the central regions are
the most extreme ones that one can expect within the WFRMS scenario. Indeed
some dilution of this processed material is expected in the radiative
envelope of the polluters (see \S~4) and then eventually later with the
intracluster gas (see \S~6). We are thus looking for stronger abundance
variations in the stellar core than the presently observed ones for the
elements involved in the CNO, NeNa and MgAl cycles.
  
We first investigate the nucleosynthesis that occurs within the convective
core of a main sequence 60~\Ms{} star and we explore in detail 
the uncertainties of the nuclear reaction rates. The evolution of the
central abundances of the key elements is presented in Fig.~\ref{fig:AbC}
as a function of the central abundance of hydrogen for the various sets of
nuclear reactions (see Table~\ref{tab:NR}).

\subsection{CNO nucleosynthesis}

In the 60~\Ms{} star the central temperature on the main sequence 
varies from $48 \times 10^6$~K to $75 \times 10^6$~K (see Fig.~\ref{fig:Tc}).
The CNO cycle thus rapidly reaches equilibrium at the beginning of
H-burning. As a result the central abundance of \el{C}{12} drops suddenly
by about an order of magnitude, while that of \el{N}{14} increases by a
factor of 29. Slightly later \el{O}{16} does reach its equilibrium value,
which is 1.6~dex below its initial abundance. After this adjustment phase
at the very beginning of the main sequence, the CNO elements stay at their
equilibrium level.  These predictions are identical for all the 60~\Ms{}
models presented here because they were all computed with NACRE
prescriptions for the CNO-cycle reaction rates (no modification of these
rates has been published since the NACRE compilation was made available).

\subsection{NeNa and MgAl predictions with the NACRE nominal values - Model 60rA}

Model 60rA was computed with the NACRE nominal values for the whole nuclear
network (full lines on Fig.~\ref{fig:AbC}).

As can be seen on Fig.~\ref{fig:AbC}, the central abundance of \el{Na}{23}
shows a three steps evolution on the main sequence: a first rapid raise by 
1.2~dex, then a more progressive increase of 0.7~dex until the mass fraction 
of hydrogen at the center, X$_\text{c}$, equals 0.4, and finally a decrease.
The initial feature is due to proton-captures on
\el{Ne}{21} and \el{Ne}{22} which are the fastest reactions of the NeNa
chain (see Fig.~\ref{fig:NR}). As proton-capture on \el{Na}{23} is slower,
the abundance of sodium increases until the complete consumption of
\el{Ne}{21} and \el{Ne}{22}. Later on \el{Ne}{20} starts
burning in favor of sodium. The slow increase is due to the competition
between the burning of \el{Ne}{20} and that of \el{Na}{23} via the channels
($p,\gamma$) or ($p,\alpha$). When using NACRE nominal values 
the transformation of \el{Ne}{20} into \el{Na}{23} is more efficient than 
the \el{Na}{23} destruction for $T$ inferior to $\sim$ $50 \times 10^6$~K 
which results in a slow increase of \el{Na}{23}. When the temperature exceeds
$50 \times 10^6$~K the situation reverses and the abundance of \el{Na}{23} 
decreases.

Let us now turn our attention to Mg and Al. In model 60rA the central
abundance of \el{Mg}{24} increases slightly during the main sequence.
That of \el{Mg}{25} first decreases by nearly 4~dex, then it presents a
sawtooth behavior until X$_\text{c}$ equals $\sim$ ~0.2 and finally it 
increases.
\el{Mg}{26} first slightly increases before decreasing 
during most of the core H-burning phase and increasing again when X$_\text{c}$
becomes lower than $\sim$ 0.07. The behavior of Mg (total) follows mainly
that of \el{Mg}{24}.  Last but not least, the abundance of Al increases.

The increase of the \el{Mg}{24} abundance is due to proton-captures on
\el{Na}{23}. The very efficient destruction of $^{25}$Mg is due to the
reaction \el{Mg}{25}$(p,\gamma)$\el{Al}{26}. The sawtooth behavior results
from small instabilities affecting the size of the convective 
core.\footnote{When the size of the convective core slightly increases, some
$^{25}$Mg is dredged down. For most elements, the
quantity dredged down is small compared to the abundance in the core;
however in the case of \el{Mg}{25} which is severely depleted in the core,
this temporarily produces a small increase of the central abundance.} The
slight production of this element at the end of the main sequence is
due to the proton-captures on \el{Mg}{24}. The initial raise of the
abundance of \el{Mg}{26} results from the destruction of \el{Mg}{25}
described above.\footnote{The abundance of the unstable isotope \el{Al}{26}
  which decays into \el{Mg}{26} is included in the total \el{Mg}{26}
  abundance.} When all the \el{Mg}{25} is consumed within the core, 
\el{Mg}{26} is converted into aluminum.

Table~\ref{tab:cen} sums-up the mean variations of the central abundances
during the main sequence for the NACRE nominal reaction rates. O and Na are
respectively depleted and produced as required by the data. On the other
hand the increase of the total magnesium we obtain is at odds with the
observed Mg-Al anticorrelation. These results agree with those of
\citet{ArnouldGoriely1999} who investigated in details
hydrogen-nucleosynthesis at constant temperature and solar metallicity with
NACRE reaction rates. They found a production of Na and they argue that it
is dominated by uncertainties on the rates for $T>50\times 10^6$~K. Also
\el{Mg}{24} is not destroyed even when experimental errors on rates are
taken into account for $T<70\times 10^6$~K. Production of aluminum is due
to the burning of both \el{Mg}{25} and \el{Mg}{26}.

\subsection{NeNa and MgAl predictions with the Illiadis nominal values - Model 60rB}

In model 60rB we take into account the nominal rates given by
\citet{IliadisStarrfield2001} and
\citet{HaleChampagne2002,HaleChampagne2004} for the reactions involved in
the NeNa and MgAl chains. The differences with the NACRE prescriptions can
be seen in Fig.~\ref{fig:NR}. At $T=50 \times 10^6$~K the rates of 
the \el{Ne}{22}$(p,\gamma)$ and
\el{Na}{23}$(p,\gamma)$ reactions are lowered respectively by 1.7~dex and
1.3~dex compared to NACRE nominal values, while the rate of
\el{Na}{23}$(p,\alpha)$ is 0.6~dex higher. For \el{Al}{27}$(p,\gamma)$ the
new rate is increased by 1~dex at $T=50 \times 10^6$~K, but is unchanged
for $T>70 \times 10^6$~K. Finally the rate of the
\el{Al}{27}$(p,\alpha)$\el{Mg}{24} reaction is now higher by 0.7 to 1.1~dex
in the range of temperature of interest. The 
prescriptions for the other reactions are unchanged.

This update does not affect the global
structure and evolution of the star, as the energizing rates concern the
CNO cycle which is not modified. Thus the lifetime, the size of the 
convective core as well as the mass loss history are identical in all our 
models.

The key reaction for the NeNa chain is \el{Ne}{20}$(p,\gamma)$ as
\el{Ne}{20} is the most abundant neon isotope and its destruction by
proton-captures is much slower than that of the \el{Ne}{21} and \el{Ne}{22}
(by 2.7 and 3.7 orders of magnitude respectively). Since the rate of this
reaction is unchanged the behavior of the Ne isotopes is barely affected
with respect to the results presented in \S~3.2. On the other hand because
the rate of proton-capture on \el{Na}{23} is reduced now with respect to
the former case, the consumption of sodium is lowered (its abundance does
not decrease at the end of the main sequence). As a consequence the
abundances of \el{Mg}{24} and of the total magnesium do not increase
although one does not obtain the decrease required by the observations.
Finally the \el{Mg}{26} burning is favored so that aluminum is produced
earlier on the main sequence although its final abundance reaches the same
value as in model 60rA.

In summary the new set of nuclear reactions resolves partly the
difficulties encountered by model 60rA regarding the anticorrelation
between Mg and Al. However the predicted magnesium isotopic ratios are
still in conflict with the data by
\citet{YongGrundahl2003,YongGrundahl2005}.

\subsection{NeNa and MgAl predictions with experimental extreme values - Model 60rC}

In order to try and solve the remaining difficulties, we look now for the
most favorable set of nuclear reactions by considering the published
experimental limits. In model 60rC (set~C) we make the following
assumptions (see Table~\ref{tab:NR}): We take the lower limits for the
burning rates of all the neon isotopes in order to lower the overall
production of sodium. We also take the lower limit for the
\el{Na}{23}($p,\gamma$) reaction which becomes now the slowest reaction of
the NeNa chain. This reduces the linkage between the two chains and
disfavors the production of \el{Mg}{24}. Regarding the MgAl chains, we take
the upper limits for the destruction rates of the magnesium isotopes and
the lower limits for the burning rates of \el{Al}{27}.

The corresponding predictions are shown by the short-dashed lines in
Fig.~\ref{fig:AbC}. The overall increase of the \el{Na}{23} abundance is
more modest than in the previous models. One obtains now an important
decrease of the \el{Mg}{26} abundance, while the final predictions for
\el{Mg}{25} and \el{Al}{27} are essentially not affected with respect to
sets~A-B. Again, \el{Mg}{24} stays constant during the hydrogen-burning
phase. Nevertheless the total magnesium abundance decreases by 0.1~dex in
this case.
The stronger variation observed could come from more massive stars where 
the central temperature is hotter and the burning rate of \el{Mg}{24} 
is faster (see \S~5).

\subsection{Model 60rD}
As shown previously the difficulty regarding the Mg destruction 
comes mainly from the fact that the central temperature reaches the 
required extreme values only at the very end of the main sequence (see 
Fig.~\ref{fig:Tc}). We thus note that the use of the rate published
for the \el{Mg}{24}$(p,\gamma)$ reaction \cite{PowellIliadis1999} is not 
compatible with significant Mg depletion within the core of massive 
main sequence stars. This led us to tentatively modify this rate 
in order to reconcile the theoretical predictions with the abundance data.

In model 60rD we keep all the reaction rates as in 60rC
except for the burning rate of \el{Mg}{24} that we artificially
  enhance by a factor of 10$^3$ and 10$^{1.5}$ at $50$ and
  $60 \times10^6$~K respectively (the recommended rate is used for temperature 
  lower than $40\times 10^6$~K and higher than $70\times 10^6$~K).
With these assumptions this reaction becomes as efficient as the other ones
involved in the MgAl chain (see the arrows on Fig.~\ref{fig:NR}).

As a result (see the dotted lines in Fig.~\ref{fig:NR}) the \el{Mg}{24}
abundance decreases by about 0.4~dex when the central hydrogen mass
fraction equals $\sim$ 0.2. At the end of the main sequence \el{Mg}{25} and
\el{Mg}{26} are respectively slightly destroyed and produced.
Simultaneously the total magnesium abundance decreases by 0.3~dex
while that of \el{Al}{27} increases by 1.2~dex for a central
hydrogen mass of 0.2. These predictions are in good agreement with
  the observational constraints.

\subsection{Summary} 

Our models were computed with the metallicity of NGC~6752 ([Fe/H]=-1.5)
with the aim of comparing our theoretical predictions with the observations
in this GC. In this section we have focused on the nucleosynthesis 
within the core of a 60~M$_{\odot}$ main sequence star. The corresponding
abundance variations are thus the most extreme ones one might expect, 
and they cannot be compared directly with the observational data. 
However we see already that we hold a serious candidate for building up 
the observed inhomogeneities and that it is worthwhile to explore further the
WFRMS scenario.

\begin{table}[htbp]
  \caption{Mean core abundance variations during MS of the
    60~\Ms{} models. $\del{X}$ refers to the amplitude (in dex) of variation of
    element X. For NGC~6752 it refers to the amplitude variation observed 
    between polluted and unpolluted stars }
  \label{tab:cen}
  \begin{tabular}{ccccccc}
    \hline
    Model & $\del{C}$ & $\del{N}$ &$\del{O}$ &$\del{Na}$ &$\del{Mg}$
    &$\del{Al}$  \\
    \hline
    \hline
    NGC~6752 & -0.7 & 1.7 & -1.0 & 0.9 & -0.3 & 1.4 \\
    \hline
    60rA     & -0.9 & 1.5 & -1.8 & 1.6 & 0.5  & 0.7 \\
    60rB     & -0.9 & 1.5 & -1.8 & 1.6 & 0.0  & 0.7 \\
    60rC     & -0.9 & 1.5 & -1.8 & 1.0 & -0.1 & 0.7 \\
    60rD     & -0.9 & 1.5 & -1.8 & 1.0 & -0.4 & 1.3 \\
    \hline
  \end{tabular}
\end{table}

The situation can be summarized by a close inspection of
  Table~\ref{tab:cen} where we list the most extreme abundance variations
  between polluted and unpolluted stars in NGC~6752 as well as in the core
  of our 60~M$_{\odot}$ models. We see that the comparison is
  rather favorable as far as the O-Na anticorrelation is concerned and
that sets~A and~B lead to higher Na production than set~C. On the other
hand, the Mg-Al anticorrelation appears to be more difficult to reproduce.
The use of sets A and B lead to a Mg-Al correlation, while that of set C
builds up a weak Mg-Al anticorrelation. The situation becomes however more
favorable when one increases the rate of the
\el{Mg}{24}$(p,\gamma)$\el{Al}{25} reaction as in model 60rD.  In
that case \el{Mg}{24} is destroyed while the abundance of \el{Mg}{25} is
barely affected and that of \el{Mg}{26} slightly increases. This is in
agreement with the observational constraints obtained on the Mg isotopes in
NGC~6752 stars by \citet{YongGrundahl2005}.

Let us note that the H-burning signatures we describe are similar to those
found in the H-burning shell during central He-burning. Indeed in this
shell the temperature does not surpass $60 \times 10^6$~K. Also the
situation is not very different in rotating and non rotating models, the
central temperatures and densities being weakly affected by rotation.

This study emphasizes the importance of some reactions: proton-captures on
\el{Ne}{20} and \el{Na}{23} respectively govern the amount of sodium
produced and the linkage between the NeNa and MgAl chains. Concerning the
MgAl chain, the key reaction is the proton-capture on \el{Mg}{24}. 
A raise of its rate with respect to the published values is 
required to reproduce the extreme variations of magnesium and aluminum in
the central region of a 60~\Ms{} star.  This is due to the fact that the
required temperature (of about 72-78 $\times 10^6$K, see Prantzos \&
Charbonnel in preparation) is reached in the stellar core only at the very
end of the main sequence in this object (see Fig.~\ref{fig:Tc}).

\section{Mixing and ejection of matter in a rotating 60~\Ms{} star}

In the previous section we discussed what happens from the nucleosynthetic
point of view in the core of a (rotating or non-rotating) 60~\Ms{} star.
We will now describe the critical effects of rotation on stellar mass loss
and internal chemical structure. Finally we will follow the evolution of
the surface abundances and compare our predictions with the data in GC
stars.

\subsection{Mass loss}

\begin{figure}[htbp]
  \includegraphics[width=0.50\textwidth]{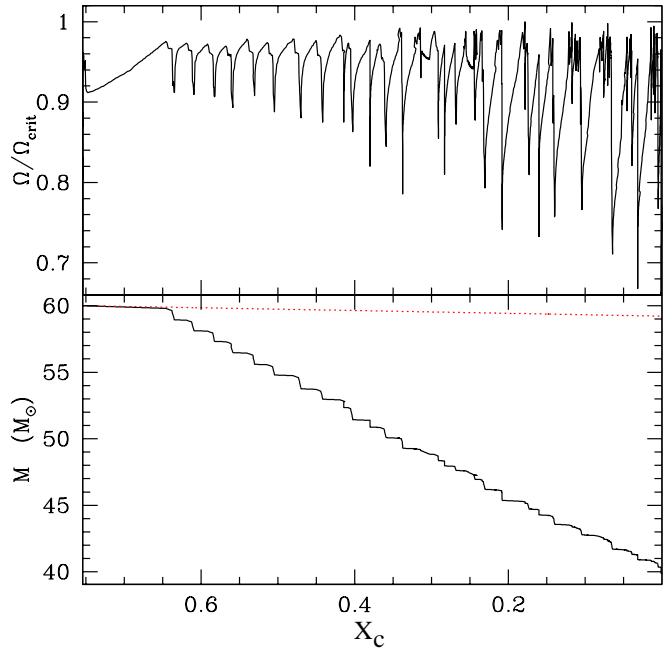}
  \caption{(top) Ratio of the surface rotation velocity to the break-up value 
    for model 60rC. (bottom) Evolution of the total stellar mass for models
    60rC and 60sC (full and dotted lines respectively) along the main
    sequence (the abscissa is the central mass fraction of hydrogen)  }
  \label{fig:crit}
\end{figure}

\begin{figure*}[tbp]
  \includegraphics[width=0.50\textwidth]{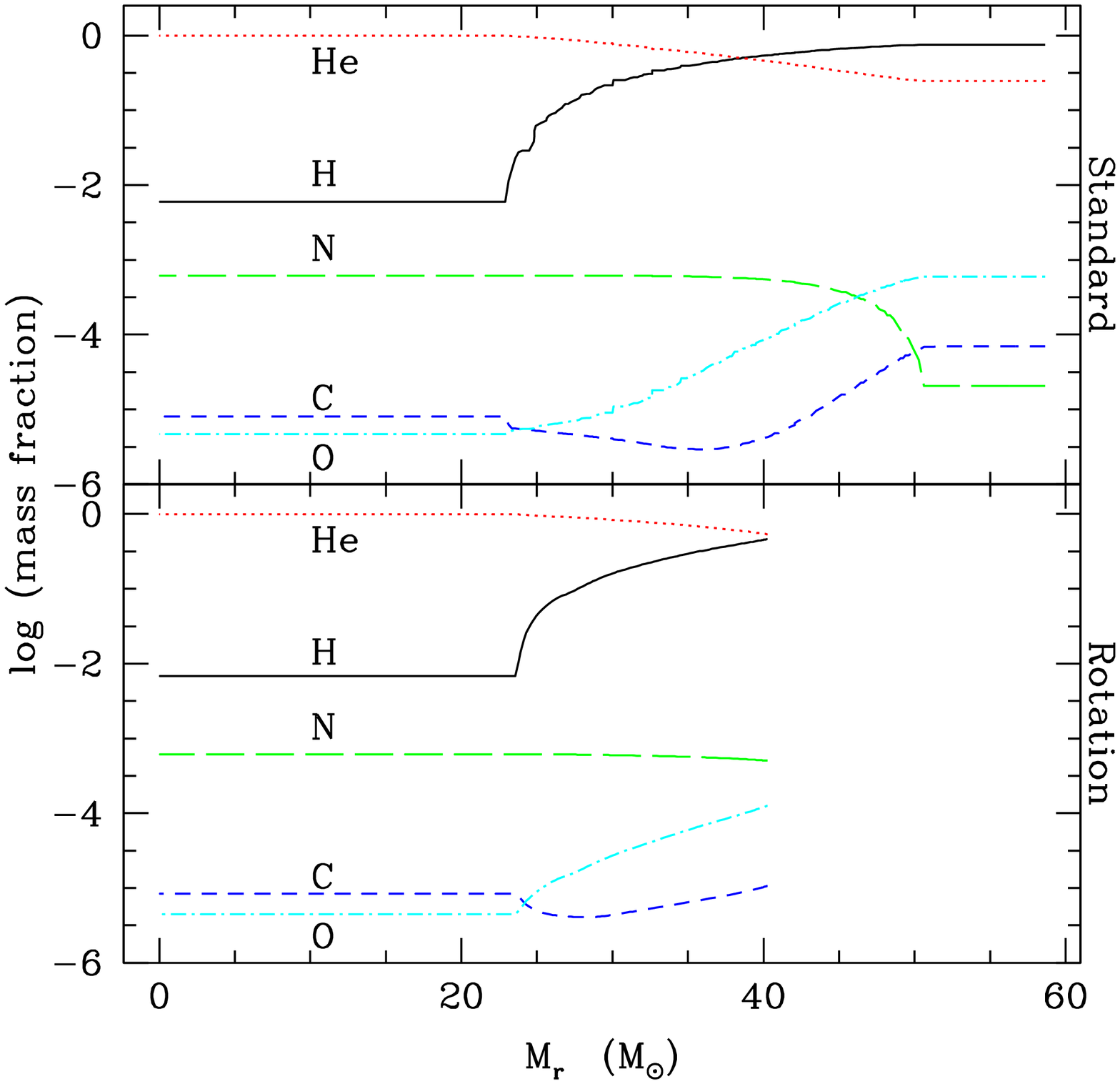}
  \includegraphics[width=0.50\textwidth]{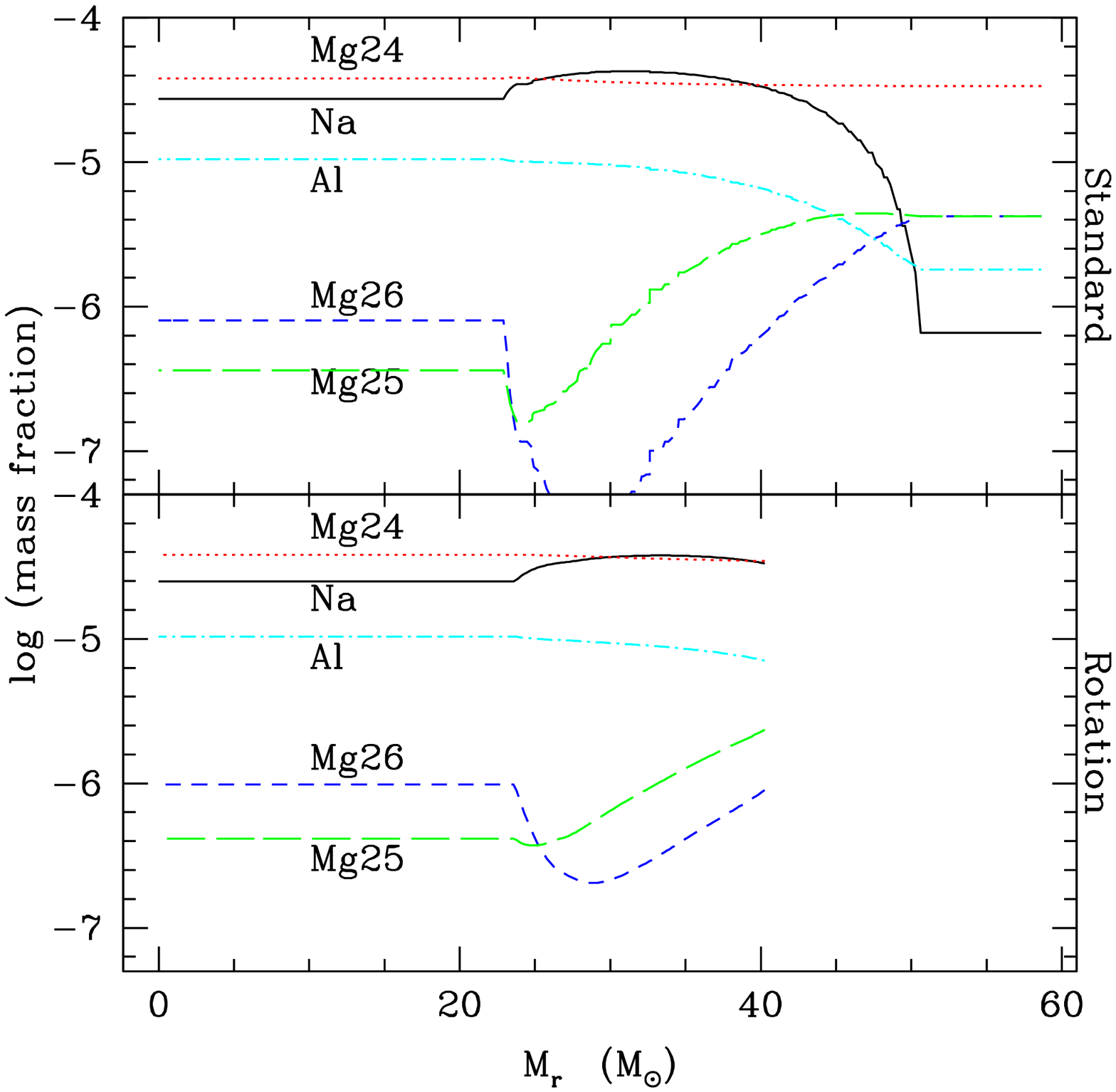}
  \caption{Abundance profiles at the end of main sequence
    ($X_c \simeq 0.01$) for the 60~\Ms{} models in the standard case (model
    60sC, top) and for $V_\text{ini} =$ 800~km~s$^{-1}$ (model 60rC,
    bottom) }
  \label{fig:Prof}
\end{figure*}

Fig.~\ref{fig:crit} shows the evolution with time of the ratio between the
surface and the break-up velocity in our 60~\Ms{} model with $V_\text{ini}
= 800$~km~s$^{-1}$ .  This ratio lies around 0.95 at the beginning
of the main sequence.  Rapidly, it falls down to 0.91 due to the
establishment of differential rotation in the star: during this
initial phase angular momentum is carried from the external
regions into the internal layers and the core accelerates (see
\citealt{MeynetMaeder2000}).  Later on $\Omega/\Omega_\text{crit}$
increases up to reach unity (see \S~2).  However we force the model
to stay below this level in order to avoid numerical difficulties (see
\S~2.2). Fig.~\ref{fig:crit} shows that all the episodes of strong mass
loss correspond to a decrease of the surface velocity below the critical
value. Then surface rotation velocity raises again. This process probably
leads to the formation of a circumstellar disk.  We make the hypothesis
that this material is eventually lost by the star.

During the whole main sequence mass loss is dominated by this strong
rotation-induced mechanical wind and is increased by more than a factor of
24 with respect to the standard case (e.g., model 60rC loses
20~\Ms{} whereas the non rotating model expels less than 1~\Ms).

As we just described, fast rotation does change the stellar mass loss 
{\sl{quantitatively}}. But it has also a crucial {\sl{qualitative}} impact on 
the properties and the topology of the ejecta. Indeed when the star rotates 
close to or at break-up the centrifugal force balances gravity. 
As a result, the mass loss is at least partly mechanically-driven, 
and the equatorial matter is
released into a Keplerian disk. Very likely an equatorial disk forms as
observed around Be stars (e.g., \citealt{PorterRivinius2003}). This ejected
material will thus be very easily retained within the GC potential well.
This is very different from the non-rotating situation where the radiative
winds escape at high velocity. This is one of the key points of our scenario
which will be discussed in detail in \S~7.

As explained in \S~2.2 the model evolves away from the
  $\Omega$--Limit when it leaves the MS. At that moment however the most
  massive stars ($M \ge 60$ M$_\odot$) encounter the $\Omega
  \Gamma$--Limit. In the present work, we suppose that the matter lost at
  the $\Omega \Gamma$--Limit is released in an equatorial disk, as in the
  case of the material lost at the $\Omega$--Limit.  When the star moves
  away from the $\Omega \Gamma$--Limit due to heavy mass loss, the
  radiatively-driven fast winds take over.

\subsection{Abundance profiles}

Since mass loss is very efficient in the rotating case, the star peels off
revealing the H-processed layers at the surface. Rotational mixing
strengthens the modifications of the surface abundances as the products of
central burning are transported outwards by various instabilities
induced by rotation.\footnote{The most important ones are the meridional
  currents and the shear turbulence (see the review by
  \citealt{Talon2004}). The inclusion of these transport mechanisms has
  improved the massive star models in many respects (see
  \citealt{MaederMeynet2000} for references); in particular, rotating models
  can reproduce the chemical enrichment observed at the surface of OBA
  stars \citep{HegerLanger2000,MeynetMaeder2000}.}

In Fig.~\ref{fig:Prof} we show the abundance profiles of various chemical
elements as a function of the Lagrangian mass for the 60~\Ms{} star
(computed using set~C for the nuclear network) at the end of the main
sequence in the standard and rotating models ($V_\text{ini}=$ 0 and
  800~km~s$^{-1}$ respectively). In both cases the size of the convective
core is very similar (respectively 23.7 and 23.5~\Ms{} in the standard and
rotating models at the end of the main sequence) as well as the final
central abundances (the central temperature differs by less than 1\%
between the two models).

Outside the convective core however the abundance profiles show some
notable differences. The most striking effect is due to mass loss which is
strongly enhanced in the rotating model. The induced strong ``peeling''
exposes the nuclear regions. As a consequence the surface abundances and
wind at the end of the main sequence hold the signatures of H-processing.

\subsection{Composition of the ejecta}

\begin{figure}[htbp]
  \includegraphics[width=0.48\textwidth]{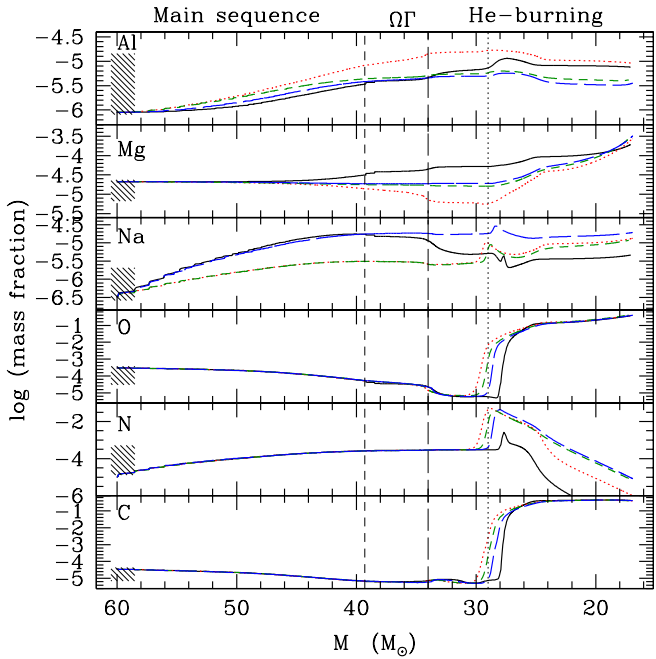}
  \caption{Evolution of the surface abundances as a function of remaining
    mass for the 60~\Ms{} rotating models computed with various sets of
    nuclear reactions : set~A (full line), set~B (long dashed lines), set~C
    (short dashed lines) and set~D (dotted lines). The short-dashed
    vertical line indicates the end of the main sequence. The $\Omega
    \Gamma$--Limit phase is located between the short- and long-dashed
    lines. The dotted line marks the moment when He-burning products start
    to appear at the surface. The shaded boxes on the left indicate the
    amplitudes of the abundance variations observed at the surface of
    low-mass stars in NGC~6752.}
  \label{fig:AbSurf}
\end{figure}

During the evolution of our 60~\Ms{} rotating models we can distinguish the
following phases:
\begin{enumerate}
\item The very beginning of the main sequence during which the surface
  velocity approaches the critical limit (between X$_\text{c} \sim
    0.76$ to 0.65 in Fig.~\ref{fig:crit}).
\item The rest of the main sequence when the surface velocity is at
  (or very near) the critical limit (for X$_\text{c}$ below 0.65 in
    Fig.~\ref{fig:crit}).
\item The beginning of the core He-burning phase when the stellar
  luminosity is close to the Eddington value and the surface velocity stays
  close to the $\Omega \Gamma$--Limit.
\item The end of the He-burning phase. From this moment on the star 
  is away from the $\Omega \Gamma$--Limit and the mass loss is
  mainly radiatively-driven. 
\end{enumerate}
During the phases 2 and 3 described above, mass loss is mainly 
mechanically-driven and the stellar winds are supposed to be slow; 
they are enriched in H-burning products only. When the 
phase 4 starts  the stellar surface and the ejecta are enriched in both 
H-burning and He-burning products, the winds are fast.

Fig.~\ref{fig:AbSurf} presents the evolution of the surface abundances in
the rotating 60~\Ms{} models as a function of the remaining stellar mass
from the zero age main sequence up to the end of central He-burning; the
different lines correspond to models computed with the various sets of
nuclear reactions described in \S~2.1 (60rA, B, C and D). We also
  indicate the range of abundances exhibited by the low-mass stars in
  NGC~6752 (shaded boxes).

The surface abundance variations do mimic the central ones (see
Fig.~\ref{fig:AbC} and the discussion in \S~3) with some delay due to
non-instantaneous rotational mixing. In all the cases the surface
abundances of carbon and oxygen at the end of the main sequence fall
respectively by 0.9 and 1~dex with respect to their original values,
whereas that of nitrogen raises by 1.5~dex. The sodium surface abundance
increases by 0.8 or 1.6~dex depending on the assumed nuclear reaction
rates. Aluminum increases in all the cases but in a more efficient way when
one enforces \el{Mg}{24}-burning by increasing significantly the
corresponding reaction rate (set D). In summary and as discussed
previously, the O-Na anticorrelation naturally shows up in all the cases.
However one has to call for an increase of the $^{24}$Mg(p,$\gamma$)
reaction rate in order to correctly build up the Mg-Al anticorrelation.

The H-burning signatures are reinforced at the stellar surface at the
beginning of the central He-burning phase mainly as a result of previous
mass loss episodes.

However later on the products of He-burning start to show up at the
surface. In particular the oxygen abundance raises steeply from $10^{-5}$
to nearly 0.6 in mass fraction when the total stellar mass is lower than
$\sim$ 28~\Ms. C and N also display strong enhancement at this phase.
After the abundance of nitrogen decreases being transformed into
  \el{Ne}{22}.

In summary the ejecta of fast rotating massive stars display the chemical
patterns observed in GC stars both on the main sequence and during the
$\Omega \Gamma$--Limit
phase.  Importantly, these are exactly the phases where matter is ejected
by gently blowing winds which can be easily retained within the GC
potential well. At the opposite, the He-burning products are released in
the interstellar matter through fast radiatively-driven winds. We will
come back to that crucial point of the WFRMS scenario in the discussion.

\subsection{Effect of the initial rotation velocity}

In order to investigate the impact of the initial rotation velocity on the
overall predictions we have computed a 60~\Ms{} model with an initial
velocity of 600~km~s$^{-1}$ (model 60rE, with set C) instead of
800~km~s$^{-1}$ (model 60rC). The mixing of chemicals is less efficient in
the slowest model as the shear turbulence, which dominates the transport
processes, decreases with lower angular velocity. However
model 60rE reaches the break-up velocity later during the main sequence
(when the central hydrogen mass fraction is 0.4 instead of 0.65 in 
model 60rC). Both effects tend to compensate each other: the
mixing has a smaller efficiency but has more time to act before the star
reaches the break-up. This leads to abundance variations which are even
more pronounced. As an example the nitrogen abundance in winds when
break-up is reached is 55\% higher in model 60rE than in model 60rC (see
Table~\ref{tab:yields}).

When the stars are at break-up, the mass loss rate is higher for a lower
initial rotation velocity: compared to the standard (non-rotating) case it
is 29 and 24 times higher in models 60rE and 60rC respectively. The total
stellar mass at the end of the main sequence is higher (by $\sim$ 3~\Ms{})
in model 60rC. These differences play only a minor role for the composition
of the yields which differ by only a few percents between the two models.

Let us note finally that with an initial velocity around 300~km~s$^{-1}$
the 60~\Ms{} star would fail to reach the break-up and the conditions
for the WFRMS scenario would not be fulfilled. If we infer models with
various initial mass and metallicity (Ekstr\"om et al, in preparation), the
60~\Ms{} would reach the critical velocity with initial velocity as low as
400~km~s$^{-1}$. However fast rotation is possibly a characteristic of
massive stars in a very dense environment such as forming GC where
multiplicity and stellar collisions can play an important role. This point
is discussed in more detail in \S~7.

\subsection{Summary}

To sum-up, the WFRMS appears to be very promising. Rotational mixing
efficiently transports elements from the convective core to the surface. A
high initial rotation velocity allows the star to reach break-up early on
the main sequence and to eject important quantities of material loaded with
H-burning products. This material is probably ejected through a
slow wind and has great chance to remain in the GC potential well (see
\S~7).

The abundance patterns at the stellar surface and in the ejecta follow
those created in the core with some delay. The amplitude of the predicted
O-Na anticorrelation reproduces well the observational feature. Although
the 60~\Ms{} star produces too few Al when one sticks to the nuclear
experimental value for the \el{Mg}{24}(p,$\gamma$) reaction rate, 
an increase of this rate allows one to explain the observed Mg-Al
anticorrelation.

\section{Dependence on the initial stellar mass}

Let us now discuss how the theoretical predictions depend on the initial
stellar mass.  Standard and rotating models with initial masses ranging
between 20 and 120~\Ms{} are presented; in all the cases we used the set~C
for the nuclear reaction rates.

\begin{figure}[tbp]
  \includegraphics[width=0.50\textwidth]{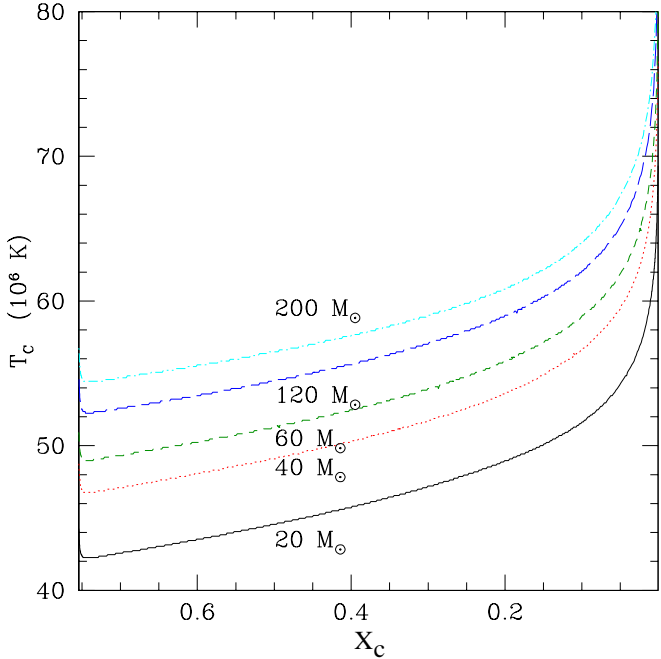}
  \caption{Evolution of the central temperature for the 20, 40, 60,
    120 and 200~\Ms{} rotating stars as a function of the central hydrogen
    abundance during the main sequence.}
  \label{fig:Tc}
\end{figure}

\begin{figure}[tbp]
  \includegraphics[width=0.50\textwidth]{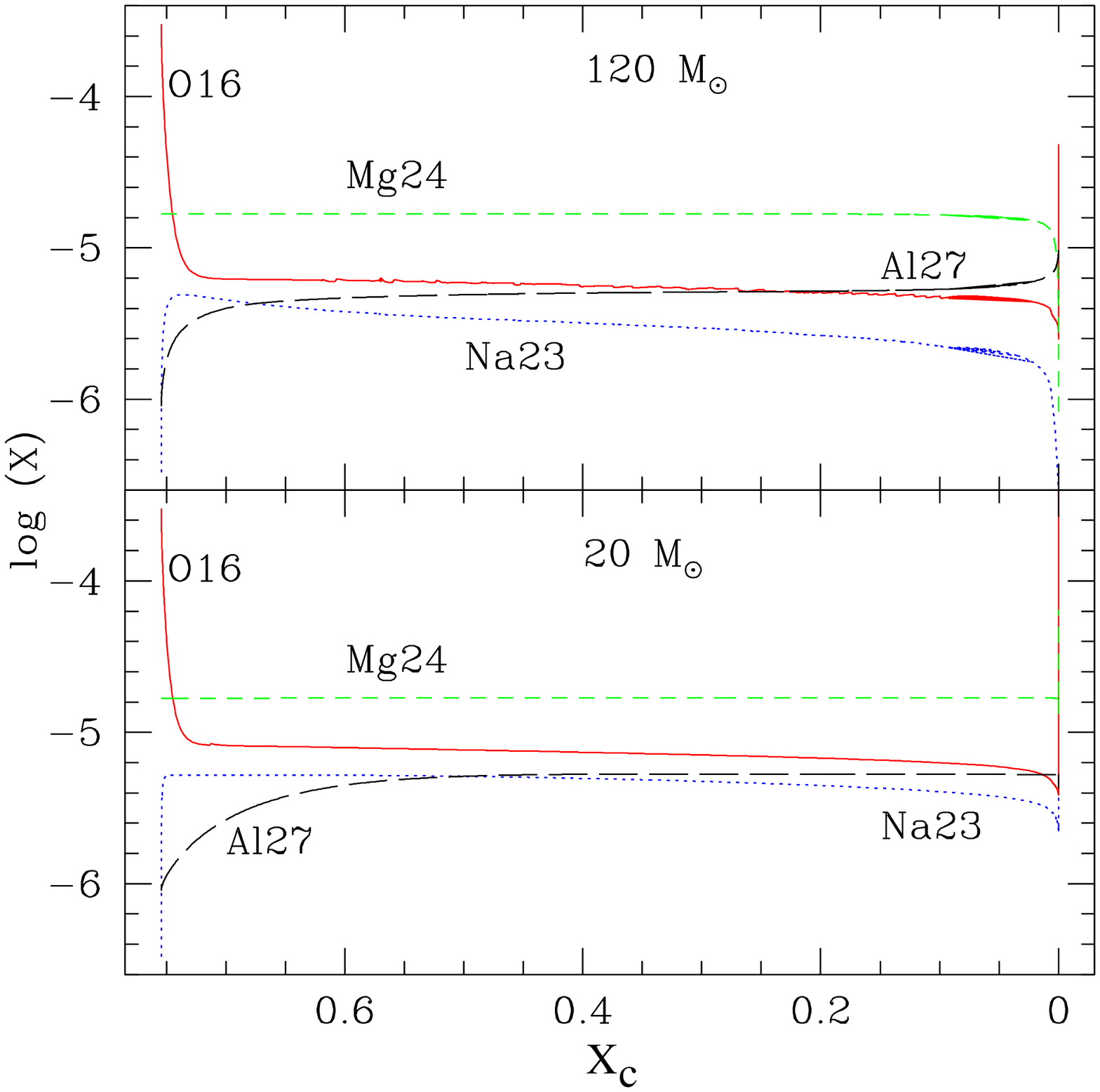}
  \caption{Evolution of the central abundances of \el{O}{16}, \el{Na}{23},
    \el{Mg}{24} and \el{Al}{27} as a function of the central hydrogen
    abundance on the main sequence for the 20 and 120~\Ms{} rotating models.}
  \label{fig:cen}
\end{figure}

\begin{figure}[t]
  \includegraphics[width=0.5\textwidth]{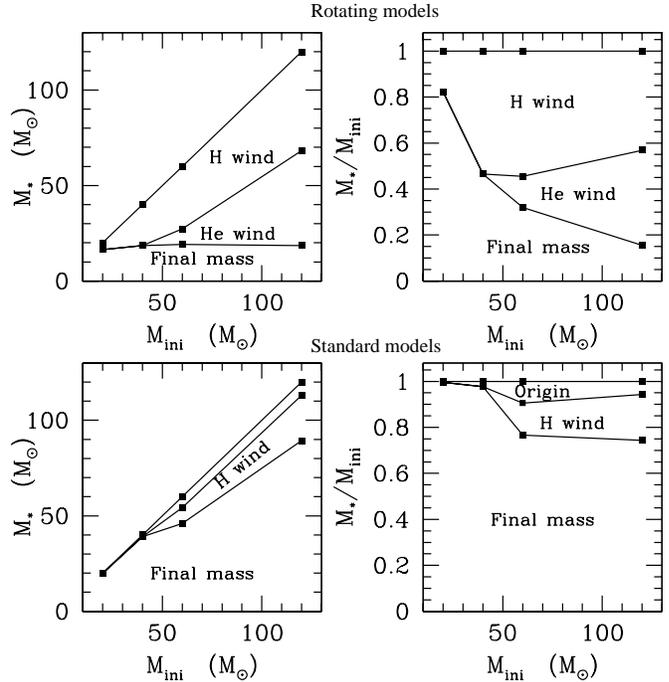}
  \caption{Composition of the matter ejected in the winds as function of 
    initial stellar mass for the rotating and standard models (upper and
    lower panels respectively).  The ejecta are presented in solar mass
    (left) or scaled with respect to initial mass (right).  The winds
    enriched in H- and He-burning products are quoted respectively as ``H
    wind'' and ``He wind''. ``Origin'' refers to matter which is not
    modified by nuclear burning. ``Final mass'' indicates the mass of the
    stars at the end of central He-burning.}
  \label{fig:Ejec}
\end{figure}

\subsection{Nucleosynthesis and mixing}

The initial mass has a direct effect on nucleosynthesis through the changes
in the central temperature (see Fig.~\ref{fig:Tc}).  As a result the NeNa
and MgAl chains are more active in the warmer convective core of more
massive stars.

The evolution along the main sequence of the central abundances of some
interesting nuclei is shown in Fig.~\ref{fig:cen} for the 20 and 120~\Ms.
It can be compared to Fig.~\ref{fig:AbC}. The following differences can be
noted:
\begin{itemize}
\item The CNO equilibrium value of O is slightly lower in the higher
  temperature regime of the 120~\Ms{} stellar model.
\item \el{Na}{23} is first produced to the same extend in both models, but
  later on it decreases faster in the more massive star.
\item In the 120~\Ms{} model \el{Mg}{24} decreases by 1.2~dex at the very
  end of the H-burning phase, while the 20~\Ms{} never reaches sufficiently
  high temperatures for this element to burn.
\item \el{Al}{27} is produced earlier in the 120~\Ms{} model. Nevertheless
  the plateau stays at the same level in both models. Only at the very end
  of the main sequence, does the Al abundance rapidly increase in the more
  massive star as a result of $^{24}$Mg burning.
\end{itemize}

Thus the main relevant difference between both stars concerns the MgAl
chain. Only in the 120~\Ms{} model and at the very end of central H-burning
are \el{Mg}{24} and \el{Al}{27} respectively destroyed and produced.  Even
the 200~\Ms{} star does not reach central temperature high enough to
convert \el{Mg}{24} before the very end of main sequence.

As discussed in \citet{MeynetMaeder2000}, rotational mixing is more
efficient when the initial stellar mass is higher, favoring the transport
of the nuclear products outwards. At a given evolutionary stage, more
massive stars also present stronger winds, both radiatively- and
mechanically-induced.

\begin{figure}[htbp]
  \includegraphics[width=0.50\textwidth]{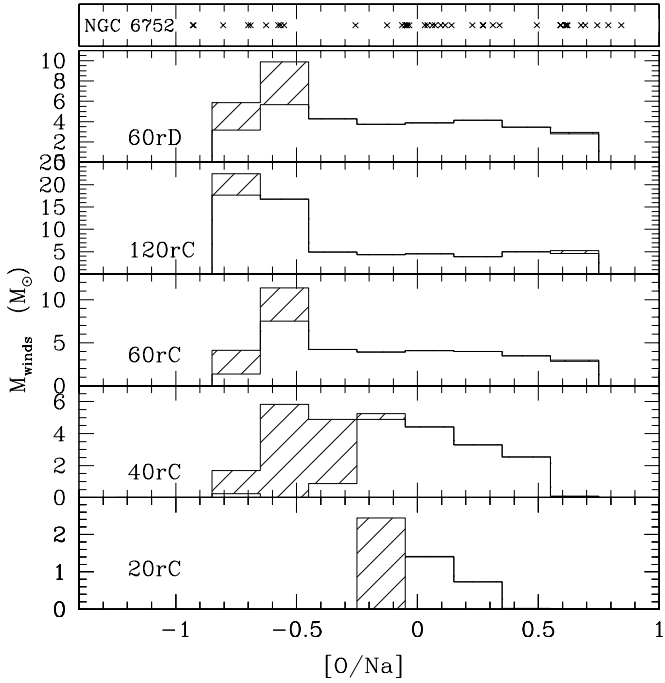}
  \caption{    
    (Top) Crosses are observed [O/Na] ratio in individual NGC~6752 stars computed from
    the data of \citet{GrundahlBriley2002} and \citet{YongGrundahl2003}.
    (Other panels) Theoretical histograms built from the wind composition
    of rotating massive stars \modif{in which 30\% of pristine matter is added
    as required by the Li behavior (see the text)}. 
    The white and shaded areas indicate respectively the slow (i.e., when 
    the star is at breakup) and fast winds}
  \label{fig:obsONa}
\end{figure}

\begin{figure}[htbp]
  \includegraphics[width=0.50\textwidth]{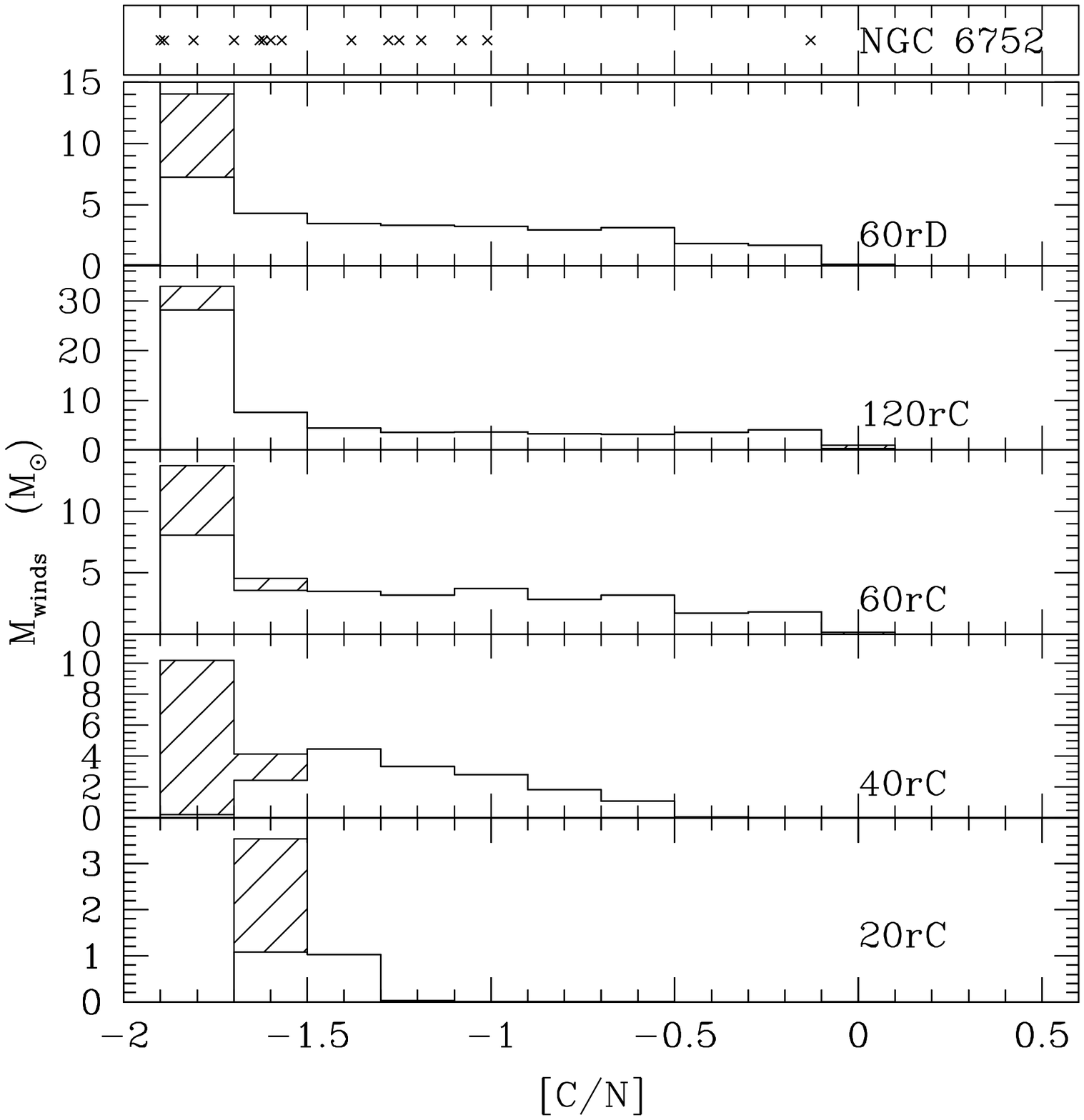}
  \caption{Same as Fig.~\ref{fig:obsONa} for [C/N] ratio (data from \citealt{CarrettaGratton2005})}
  \label{fig:obsCN}
\end{figure}

\begin{figure}[htbp]
  \includegraphics[width=0.50\textwidth]{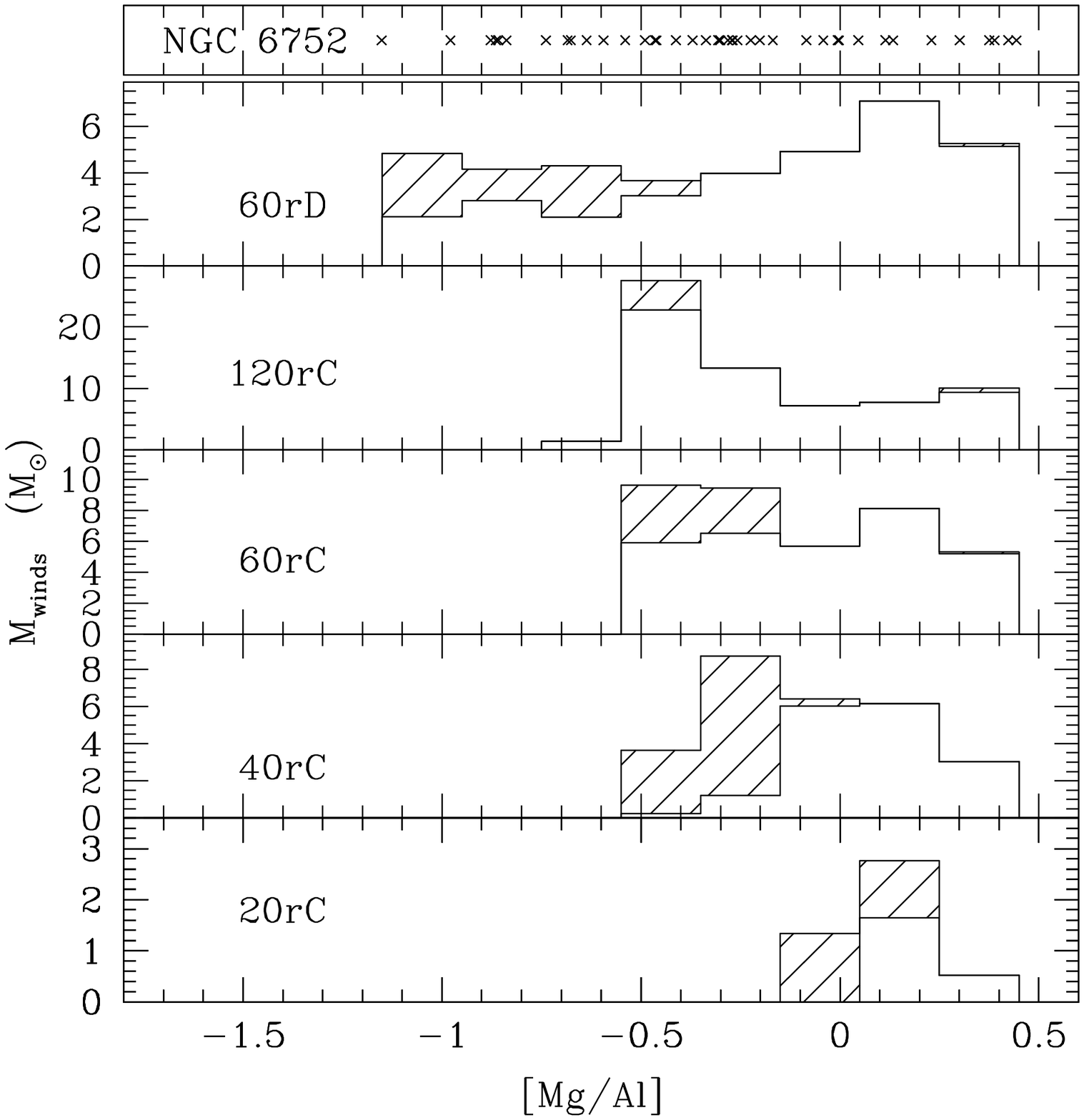}
  \caption{Same as Fig.~\ref{fig:obsONa} for [Mg/Al] ratio (data from \citealt{GrundahlBriley2002,YongGrundahl2003})}
  \label{fig:obsMgAl}
\end{figure}

\subsection{Winds}

Fig.~\ref{fig:Ejec} illustrates the differences between the standard and
rotating models for the wind composition and the remaining stellar mass at
the end of central He-burning. As already discussed, the rotating models lose 
much more mass than the standard ones due to three reasons: first, they do
undergo mechanical winds when they reach the break-up velocity; second,
their radiatively-driven winds are enhanced by the correcting factor due to
rotation; third, they enter the WR phase earlier in their evolution.

The rotating models with an initial mass above 40~\Ms{} lose about 
half of their mass through winds loaded in H-burning products. 
Moreover this matter is released very smoothly in the interstellar medium, 
with velocities low enough to remain in the GC.  This is not the case of the 
winds of non-rotating massive stars. Indeed in that case, the wind velocities 
on the main sequence are of the order of 1000 to 2000~km~s$^{-1}$ 
\citep[see e.g.][]{LamersSnow1995} and this material is probably thrown 
off the gravitational potential well of the cluster.

\modif{
\section{Comparison with observed abundance variations in GCs}

Here we explore how the ejecta of massive rotating stars can be used to
form low-mass stars displaying abundance variations in light elements.  It
is out of the scope of the present paper to present a detailed study of the
interaction between the disk enclosing fast rotating massive stars and the
surrounding protocluster gas. We will thus only make some simple
assumptions to verify if massive stars can be the progenitors of the
long-lived stars we observe today.

As the disks may be mixed with some pristine gas, we need first to evaluate
the amount of this dilution. For that we use abundance variations in Li
measured by \citet{PasquiniBonifacio2005} in NGC~6752: Li is found to be as
low as $A(Li) = 1.93$ (Alonso temperature scale) in the most polluted
stars. In the present models we do not follow explicitly the
nucleosynthesis of lithium but we can reasonably assume that this fragile
element is completely destroyed due to high temperature is the stellar
interior and to the strong mixing induced by rotation.  If we assume that
the initial lithium abundance of the intracluster gas corresponds to the
cosmological value of 2.61 \citep{CocVangioniFlam2004} and that no lithium
is present in the wind of massive stars, we need to add about 30\% of
pristine gas to those winds in order to find the extreme value of lithium
(see also Prantzos \& Charbonnel in preparation).

As a first estimate we use this dilution factor between the stellar winds
and the original matter and check whether it allows to reproduce all the
observed abundance variations. In real proto-globular cluster there will
surely be a dispersion in the amount of dilution which in turn creates a
variation in the abundances of low-mass stars.

Fig.~\ref{fig:obsONa}, \ref{fig:obsCN} and \ref{fig:obsMgAl} display the
composition of the mixed matter formed by adding 30\% of pristine gas to
massive star ejecta as well the observed abundance distributions in
NGC~6752 \citep{GrundahlBriley2002,YongGrundahl2003,CarrettaGratton2005}.
The theoretical histograms indicate the mass ejected by winds mixed with
pristine gas resulting on a given chemical composition. The white areas
correspond to the mass lost at break-up, while the hatched regions indicate
that the corresponding winds are released when stellar rotation is well
below its critical value (this occurs at the very beginning of the main
sequence and during the He-burning phase).} This latter component has thus
a high velocity and probably escapes the GC potential well. Only the H-rich
loaded winds are shown here. The He-rich winds are released at high
velocity and are supposed to escape the potential well of the GC.

Let us first concentrate on the [O/Na] distribution. The abundance pattern
in the \modif{mixed matter of} the 60 and 120~\Ms{} models covers the
entire observational range. The case of the less massive stars is
different. At first glance the predictions for the 40~\Ms{} model also
account well for the observed dispersion. This is not the case when one
considers only the slow winds ejected at break-up. The 20~\Ms{} star
follows the same trend but with a smaller extent. Let us note that the
highest [O/Na] ratio for both the 20 and 40~\Ms{} stars is shifted to the
left compared to more massive models.  This is due to the longer time required
to reach critical velocity at the beginning of the main sequence. When the
break-up is reached rotational mixing had more time to transport the
elements through the radiative envelope.

Turning our attention to the [C/N] distribution (Fig.~\ref{fig:obsCN}) we
find that the ejecta of rotating massive stars cover very nicely the
observational range. Again the 20~\Ms{} and 40~\Ms{}, being more mixed
before reaching break-up, do not expel matter with original composition
(see Table~\ref{tab:yields}).

Regarding the [Mg/Al] ratio we see that the winds of all the models
computed with the set~C (experimental limits) present lower abundance
variations than required by the observed distribution. In addition let us
recall that the magnesium isotopic ratios obtained in these models are at
odd with the observed ones: even in the case of the 120~\Ms{} aluminum is
built mainly from \el{Mg}{25} and \el{Mg}{26}. Only model 60rD covers the
whole observational range both in terms of Mg and Al abundances and of Mg
isotopic ratios. The same would be true for other masses computed with set
D. This difficulty remains even in the extreme case, where present day halo
stars would be naked cores of massive stars having undergone strong
evaporation at the end of the main sequence. Only at the end of main
sequence the 60 and 120~\Ms{} models manage to destroy \el{Mg}{24} by more
than 0.3~dex and to enhance \el{Al}{27}. However at that time the high
central temperature leads to a strong destruction of \el{Na}{23} which
falls near or under its initial value.

\begin{table*}[htbp]
  \caption{Sum-up of the main features of our models: initial mass,
  initial surface velocity, and for each evolutionary phases 
  the remaining mass and the mean mass fraction in the wind for
  several isotopes integrated from the zero age main sequence. 
  The 20, 40 and 120~\Ms{} models are computed with the reaction rates from
  set C. \modif{For rotating models the last division indicates abundance
    when He-burning products appear at the surface, or at the end of He-burning
  phase depending on the first case reached.}}
\label{tab:yields}
  \begin{tabular}{cccccccccccccc}
    \hline 
    $M_\text{ini}$ & $V_\text{ini}$ & $M_\text{f}$ & H1 & He4 & C12 & N14 &
    O16 & F19 & Na23 & Mg24 & Mg25 & Mg26 & Al27 \\
    \Ms & km~s$^{-1}$ & \Ms & \multicolumn{11}{c}{Mean mass fraction in winds} \\
    \hline
    \hline
    \multicolumn{14}{c}{\textsc{Standard Models}} \\
    \multicolumn{14}{c}{End of central H-burning} \\
    20 & 0 & 19.9 & 7.5e-1 & 2.4e-1 & 3.5e-5 & 1.0e-5 & 3.0e-4 &
    1.5e-8 & 3.3e-7 & 1.7e-5 & 2.1e-6 & 2.1e-6 & 9.0e-7 \\
    40 & 0 & 39.7 & 7.5e-1 & 2.4e-1 & 3.5e-5 & 1.0e-5 & 3.0e-4 &
    1.5e-8 & 3.3e-7 & 1.7e-5 & 2.1e-6 & 2.1e-6 & 9.0e-7 \\
    60 & 0 & 59.2 & 7.5e-1 & 2.4e-1 & 3.5e-5 & 1.0e-5 &
    3.0e-4 & 1.5e-8 & 3.3e-7 & 1.7e-5 & 2.1e-6 & 2.1e-6 & 9.0e-7 \\
    120 & 0 & 117.0 & 7.5e-1 & 2.4e-1 & 3.5e-5 & 1.0e-5 & 3.0e-4 &
    1.5e-8 & 3.3e-7 & 1.7e-5 & 2.1e-6 & 2.1e-6 & 9.0e-7 \\
    \multicolumn{14}{c}{End of central He-burning} \\
    20 & 0 & 19.9 & 7.5e-1 & 2.4e-1 & 3.5e-5 & 1.0e-5 & 3.0e-4 &
    1.5e-8 & 3.3e-7 & 1.7e-5 & 2.1e-6 & 2.1e-6 & 9.0e-7 \\
    40 & 0 & 39.1 & 7.5e-1 & 2.5e-1 & 3.5e-5 & 1.0e-5 & 3.0e-4 &
    1.5e-8 & 3.3e-7 & 1.7e-5 & 2.1e-6 & 2.1e-6 & 9.0e-7 \\    
    60 & 0 & 45.9 & 6.9e-1 & 3.1e-1 & 2.6e-5 & 7.9e-5 &
    2.3e-4 & 1.2e-8 & 1.2e-6 & 1.7e-5 & 1.6e-6 & 1.7e-6 & 1.8e-6 \\
    120 & 0 & 89.3 & 5.9e-1 & 4.1e-1 & 1.5e-5 & 1.7e-4 & 1.4e-4 &
    6.3e-9 & 2.0e-6 & 1.7e-5 & 9.9e-7 & 1.2e-6 & 3.0e-6 \\
    \hline
    \multicolumn{14}{c}{\textsc{Rotating Models}} \\
    \multicolumn{14}{c}{Break-up reached} \\
    20 & 600 & 20.0 & 7.5e-1 & 2.4e-1 & 1.1e-5 & 4.9e-5 & 2.9e-4 &
    1.5e-8 & 7.5e-7 & 1.7e-5 & 2.1e-6 & 2.2e-6 & 9.3e-7 \\
    40 & 800 & 40.0 & 7.5e-1 & 2.4e-1 & 3.5e-5 & 1.3e-5 & 3.0e-4 &
    1.6e-8 & 3.7e-7 & 1.7e-5 & 2.1e-6 & 2.2e-6 & 9.2e-7 \\
    60rE & 600 & 59.7 & 7.5e-1 & 2.5e-1 & 3.3e-5 & 1.7e-5 & 3.0e-4 &
    1.5e-8 & 4.2e-7 & 1.7e-5 & 2.1e-6 & 2.1e-6 & 9.2e-7 \\
    60rA & 800 & 59.9 & 7.5e-1 & 2.4e-1 & 3.5e-5 & 1.1e-5 & 3.0e-4 &
    1.6e-8 & 3.5e-7 & 1.7e-5 & 2.1e-6 & 2.1e-6 & 9.2e-7 \\
    60rB & 800 & 59.9 & 7.5e-1 & 2.5e-1 & 3.5e-5 & 1.1e-5 & 3.0e-4 &
    1.6e-8 & 3.5e-7 & 1.7e-5 & 2.1e-6 & 2.2e-6 & 9.2e-7 \\
    60rC & 800 & 59.9 & 7.5e-1 & 2.4e-1 & 3.5e-5 & 1.1e-5 & 3.0e-4 &
    1.6e-8 & 3.4e-7 & 1.7e-5 & 2.1e-6 & 2.2e-6 & 9.2e-7 \\
    60rD & 800 & 59.9 & 7.5e-1 & 2.5e-1 & 3.5e-5 & 1.1e-5 & 3.0e-4 & 
    1.6e-8 & 3.4e-7 & 1.7e-5 & 2.1e-6 & 2.2e-6 & 9.2e-7 \\
    120 & 800 & 119.5 & 7.5e-1 & 2.4e-1 & 3.5e-5 & 1.1e-5 & 3.0e-4
    & 1.6e-8 & 3.4e-7 & 1.7e-5 & 2.1e-6 & 2.1e-6 & 9.2e-7 \\
    \multicolumn{14}{c}{End of central H-burning} \\
    20 & 600 & 18.4 & 7.3e-1 & 2.7e-1 & 1.6e-6 & 8.8e-5 & 2.5e-4 &
    1.3e-8 & 1.5e-6 & 1.7e-5 & 1.8e-6 & 2.0e-6 & 1.3e-6 \\
    40 & 800 & 28.8 & 7.2e-1 & 2.8e-1 & 1.3e-5 & 9.3e-5 & 2.3e-4 &
    1.2e-8 & 1.4e-6 & 1.7e-5 & 1.7e-6 & 1.9e-6 & 1.6e-6 \\
    60rE & 600 & 42.7 & 6.7e-1 & 3.3e-1 & 1.9e-5 & 1.1e-4 & 2.1e-4 &
    1.0e-8 & 1.5e-6 & 1.7e-5 & 1.5e-6 & 1.6e-6 & 2.0e-6 \\
    60rA & 800 & 39.9 & 6.8e-1 & 3.2e-1 & 1.9e-5 & 1.2e-4 & 2.0e-4 &
    1.0e-8 & 7.1e-6 & 1.9e-5 & 1.4e-6 & 2.2e-6 & 1.5e-6 \\
    60rB & 800 & 39.9 & 6.8e-1 & 3.2e-1 & 1.9e-5 & 1.2e-4 & 2.0e-4 &
    1.0e-8 & 6.4e-6 & 1.7e-5 & 1.4e-6 & 2.0e-6 & 1.7e-6 \\
    60rC & 800 & 39.7 & 6.8e-1 & 3.2e-1 & 1.9e-5 & 1.2e-4 & 2.0e-4 &
    1.0e-8 & 1.7e-6 & 1.7e-5 & 1.4e-6 & 1.6e-6 & 2.1e-6 \\
    60rD & 800 & 39.9 & 6.8e-1 & 3.2e-1 & 1.9e-5 & 1.2e-4 & 2.0e-4 &
    1.0e-8 & 1.7e-6 & 1.6e-5 & 1.5e-6 & 1.8e-6 & 2.7e-6 \\
    120 & 800 & 76.6 & 6.1e-1 & 3.9e-1 & 1.7e-5 & 1.8e-4 & 1.3e-4
    & 6.3e-9 & 2.2e-6 & 1.7e-5 & 9.2e-7 & 1.1e-6 & 3.1e-6 \\
    \multicolumn{14}{c}{Appearence of He-burning products at the surface/End of central He-burning} \\
    20 & 600 & 16.4 & 7.0e-1 & 3.0e-1 & 1.6e-6 & 1.1e-4 &
    2.3e-4 & 1.1e-8 & 1.8e-6 & 1.7e-5 & 1.7e-6 & 1.9e-6 & 1.6e-6 \\
    40 & 800 & 18.5 & 5.7e-1 & 4.3e-1 & 9.3e-6 & 1.6e-4
    & 1.6e-4 & 8.0e-9 & 2.1e-6 & 1.7e-5 & 1.2e-6 & 1.3e-6 & 2.7e-6 \\
    60rE & 600 & 33.2 & 5.2e-1 & 4.8e-1 & 1.4e-5 & 1.7e-4 & 1.4e-4 &
    7.2e-9 & 2.0e-6 & 1.7e-5 & 1.1e-6 & 1.2e-6 & 3.0e-6 \\
    60rA & 800 & 30.8 & 5.4e-1 & 4.6e-1 & 1.5e-5 & 1.7e-4 & 1.5e-4 &
    7.2e-9 & 8.6e-6 & 2.5e-5 & 1.3e-6 & 2.0e-6 & 2.5e-6 \\
    60rB & 800 & 30.7 & 5.5e-1 & 4.5e-1 & 1.5e-5 & 1.7e-4 & 1.5e-4 &
    7.2e-9 & 1.0e-5 & 1.7e-5 & 1.1e-6 & 1.5e-6 & 2.6e-6 \\
    60rC & 800 & 30.7 & 5.5e-1 & 4.5e-1 & 1.5e-5 & 1.7e-4 & 1.5e-4 &
    7.3e-9 & 2.0e-6 & 1.7e-5 & 1.1e-6 & 1.2e-6 & 3.0e-6 \\
    60rD & 800 & 31.0 & 5.5e-1 & 4.5e-1 & 1.5e-5 & 1.7e-4 & 1.5e-4 &
    7.3e-9 & 2.0e-6 & 1.4e-5 & 1.1e-6 & 1.6e-6 & 5.5e-6 \\
    120 & 800 & 68.7 & 5.0e-1 & 5.0e-1 & 1.5e-5 & 2.0e-4 & 1.1e-4 &
    5.2e-9 & 2.2e-6 & 1.7e-5 & 8.4e-7 & 9.6e-7 & 3.5e-6 \\
    \hline
  \end{tabular}
\end{table*}

\section{Discussion}

Let us now discuss some details of the WFRMS scenario.

\subsection{Kinematics and topology of the ejected material}

\begin{figure}[htbp]
  \includegraphics[width=0.5\textwidth]{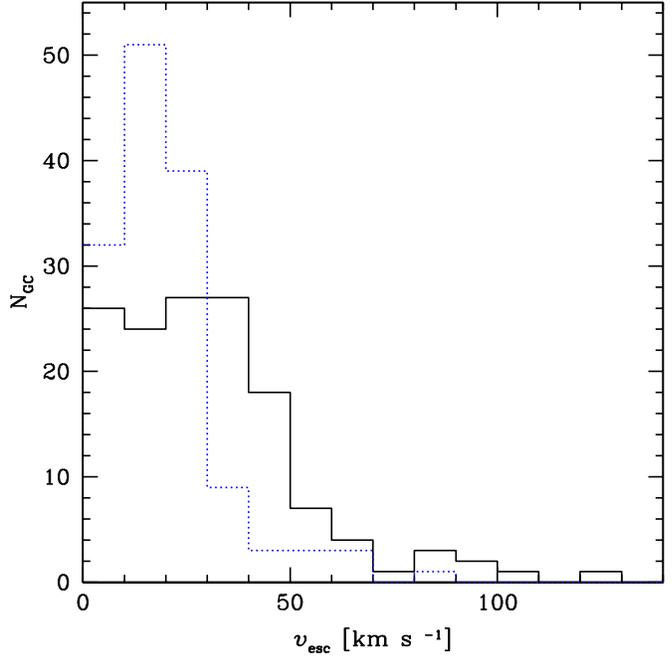}
 \caption{Histogram of the number of globular clusters having a given value
   of the escape velocity. The data are taken from \citet{GnedinZhao02}.
   The continuous line shows the results for the escape velocity at the
   center of the cluster, the dotted line at the cluster half-mass radius.}
 \label{fig:vesc}
\end{figure}

\begin{figure}[htbp]
  \includegraphics[width=0.5\textwidth]{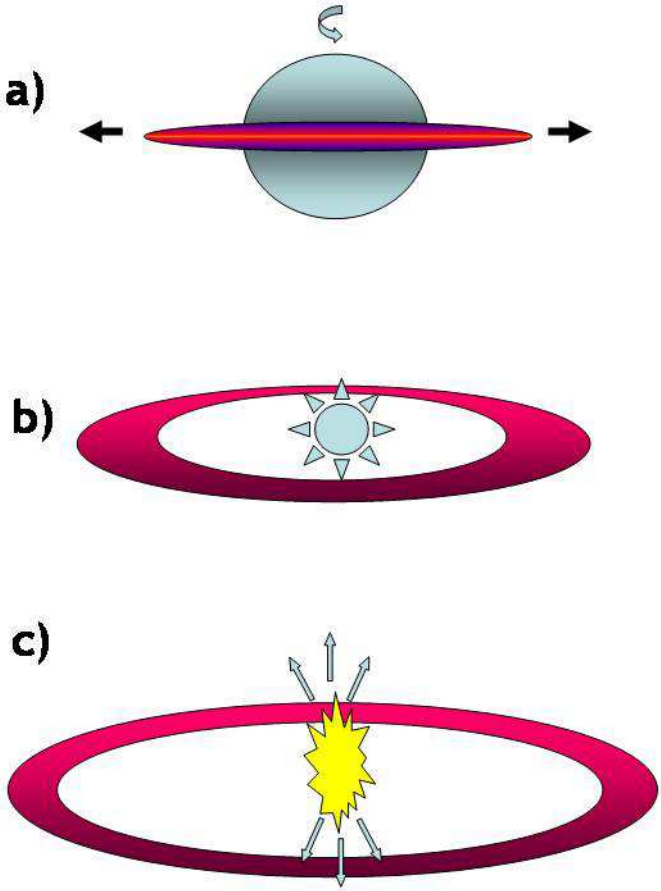}
  \caption{Schematic view of the WFRMS scenario showing the possible
    geometry of the stellar ejecta at various evolutionary phases: {\bf a)}
    during the main sequence, when the star rotates near or at the
    critical velocity, matter is preferentially ejected in the equatorial
    plane by the action of the centrifugal acceleration; {\bf b)} after the
    main sequence, the surface velocity is no long critical and the
    wind is triggered mainly by radiation. It is no long equatorial and
    becomes isotropic; {\bf c)} the supernova explosion resulting from an
    initially fast spinning star, if it occurs, may favor ejection through
    jets aligned along the rotational axis.}
 \label{fig:schema} 
\end{figure}

\begin{figure}[htbp]
  \includegraphics[width=0.5\textwidth]{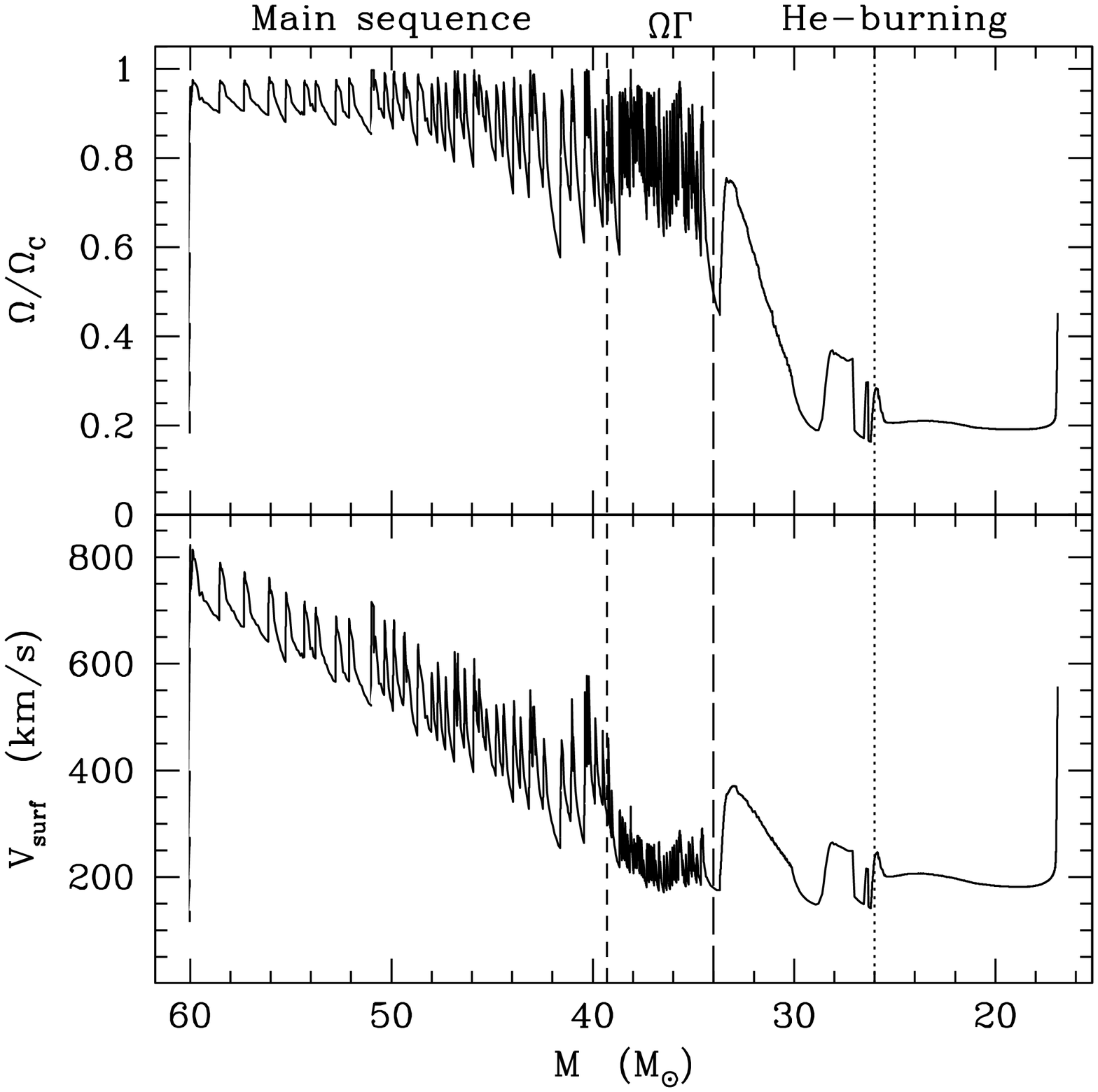}
 \caption{Evolution of the ratio between the surface angular velocity and the
   classical critical velocity (top) and of the surface equatorial velocity
   (bottom) as a function of the remaining mass for the 60~\Ms{} star.
   Vertical short-dashed line corresponds to the end of the main sequence
   and dotted line indicates the moment when He-burning products show up at
   the surface. The $\Omega \Gamma$--Limit phase is delimited by the short- and long-dashed
   lines.}
 \label{fig:Wind}
\end{figure}

Figure~\ref{fig:vesc} shows the distribution of the globular clusters as a
function of the escape velocity estimated at the center and at half--mass
radius \modif{of the present-day cluster} (continuous and dotted line 
respectively, \citealt{GnedinZhao02}). One sees that more than 60\% of the 141 
globular clusters used to build the histogram ({\it i.e.}  the large majority 
of the known galactic globular clusters) have an escape velocity at the center
between 20 and 100~km~s$^{-1}$.  The range is shifted to lower values if escape
velocities estimated at half--mass radius are used. Thus, taken at face,
these values imply that all the wind ejecta of O-type stars and of WR stars
(wind velocities of a few thousandths of km~s$^{-1}$), of Luminous Blue
Variable (wind velocities of a few hundred km~s$^{-1}$) as well as the
supernova ejecta (velocities of the ejecta of the order of
$10\,000$~km~s$^{-1}$) will be lost by the globular cluster.  This
statement must however be a little tempered by a couple of facts: first the
escape velocity of the present day globular clusters may be different from
the one at the time of their formation.  Would the present globular
clusters be the remnants of much more massive systems having lost a great
part of their initial mass, then the present day escape velocity might be a
poor estimate of the escape velocity at the time of their
formation.\footnote{More precisely we need here the escape velocity at the
  time of the chemical inhomogeneities formation, which in our present
  model would be different from the time of formation of the first stellar
  generation.} Second, the interstellar medium at the time of the globular
cluster formation was much more dense and thus the matter ejected was
slowed down by the bow shock between the ejecta and the ambient medium as
shown for instance in the numerical models by \citealt{FreyerHensler2003}
(see also \citealt{GarciaSeguraLanger1996}).  It is however difficult,
without a more precise view of the initial conditions of the globular
cluster formation to come to firm conclusion on these points.\footnote{Of
  course one could argue that we can observe young globular clusters and
  thus obtain some constraints on their formation.  However, many (if not
  all) young super star clusters observed today and believed to be young
  globular clusters, result from galaxy merging or interactions. It is not
  clear at all that the old galactic globular clusters were formed in this
  way.} In the following, in absence of a better knowledge of the globular
cluster formation conditions, we shall consider that the present escape
velocities are representative of the escape velocities at the time of the
chemical inhomogeneities formation.

Thus we have to look if our massive star models are able to eject matter at
sufficiently low velocities for allowing the ejecta to remain in the
potential well of the globular cluster. Rotation might be of great help in
this respect. Indeed, in situations where the surface equatorial velocity
is such that the centrifugal acceleration exactly counterbalances the
gravity, matter can be launched into a Keplerian orbit and an equatorial
disk easily forms as schematically shown in Fig.~\ref{fig:schema}. The
material mainly ``mechanically'' ejected into the disc will be possibly
available for further star formation, unless it falls back onto the star
and/or is ejected outside the globular cluster by further violent wind
episodes or by the supernova explosion. Let us briefly discuss these
different possibilities:
\begin{itemize}
\item {\bf Fall back of the disk material?} Be stars are main-sequence
  stars surrounded by a disc \citep{PorterRivinius2003}. The origin of the
  disk is probably stellar rotation near the critical limit as convincingly
  discussed by \citet{TownsendOwocki2004}. Thus these objects are exactly
  the observable counterparts of the stellar models we are interested in
  here. Observations (see the review by \citealt{Rivinius2005}) indicate
  that Be star disks are eroded and finally dissipate, adding their
  material to the ambient interstellar medium. Thus these observations
  would favor the lost of the disk material around Be stars and not its
  fall back onto the star.
\item {\bf Ejection by fast stellar winds?} When the star does no longer
  rotate with the critical velocity, matter is no longer preferentially
  ejected in the equatorial plane, and different situations may occur.
  First it might be that the timescale for disk dissipation is very short
  and thus that most of the disk material has already disappeared when the
  faster radiatively driven winds set in. In that case there is the
  possibility that the fast wind pushes this freshly ejected material out
  of the globular cluster, which will be then lost. The most favorable case
  for keeping the material inside the globular cluster is that the disk has
  a sufficiently long lifetime for still keeping most of its material when
  the fast winds set in. In that case, if the star is still in the blue
  part of the HR diagram (as is the case for our 60 and 120 M$_\odot$) and
  if its rotation is still sufficiently close to the critical value, the
  wind will be preferentially ejected along the rotational axis. This would
  thus prevent the winds to sweep off the disk. If the star has a
  rotational velocity well below the critical one, the wind is isotropic
  and one may hope that the fraction of the wind which may interact with
  the disk will be too small to destroy it (see Fig.~\ref{fig:schema}).
\item {\bf Will the disk be thrown away by the supernova explosion?} In
  case a black hole is formed that swallows all the mass of the
  pre-supernova, no explosion occurs and the disk will not be affected. If,
  on the other hand, a supernova explosion occurs, the fast rotation of the
  core may favor ejection along the rotational axis as in the models of
  \citet{MaedaNomoto2003}. In that case there is some chance that the disk
  will not be affected by the supernova explosion as illustrated
  schematically in Fig.~\ref{fig:schema}. \modif{Of course the disk
    material can also be destroyed and ejected out of the GC pushed away by
    the winds and/or supernova shocks from nearby massive stars. Our
    scenario will then be possible only if in relatively short timescales
    (shorter than typically the averaged time between two supernova
    events), new stars form in the disk, or more probably in the
    interaction region between the disk material and the interstellar
    medium. In that case, the fast moving material will escape following
    the nearly empty channels left behind by the star formation process, in
    line with the scenario proposed by PC06.}
\end{itemize}
These ideas are highly speculative and will have to be checked by
detailed hydrodynamical calculations in the future. For the purpose of the
present discussion, we shall suppose that the equatorial disk formed when
the star reaches the critical limit consists of material which will be made
available for forming new stars in the globular cluster.

Now, in the frame of the above hypothesis, we have to see when this occurs
during the evolution of our stellar models and whether sufficient mass with
the adequate chemical composition is ejected during these phases. The
present models show that the critical limit can be reached at two epochs:
\begin{enumerate}
\item During the main sequence, provided that the initial velocity is high
  enough. From Fig.~\ref{fig:Wind} one sees that our 60 M$_\odot$ model
  with $\upsilon_{\rm ini}=$ 800 km~s$^{-1}$ remains near the break-up
  limit during nearly the whole main sequence. At the turnoff, the star has
  lost more than 20~\Ms{}.
\item After the main sequence we saw that stars massive enough do reach the
  $\Omega\Gamma$--limit which maintains them at the break-up\footnote{At
    this stage, the critical velocity is lowered with respect to its
    classical expression (see \citealt{MaederMeynet2000}).} due to the
  raise of the luminosity. Strong mass loss ensues: During this phase the
  60~\Ms{} star expels more than 5~\Ms{} (see Fig.~\ref{fig:Wind}).  We
  suppose here that this material will still be preferentially ejected
  along the equatorial plane and will join the disk of slowly outflowing
  material.
\end{enumerate}

\subsection{Composition of the ejected material}

In the previous sections we have seen that the chemical composition of the
ejected material presents many similarities with the abundance patterns
observed at the surface of globular cluster stars. If one focuses on the
matter ejected during the main sequence, one sees however that the
theoretical variations can account only partly for the observed range in
oxygen and sodium. However slightly later in the evolution, strongly CNO
processed material is ejected and the most extreme observational cases can
be reproduced (see Table~\ref{tab:yields}). The observed variations of Mg
and Al can also be reproduced, provided the rate of the
\el{Mg}{24}$(p,\gamma)$ reaction is enhanced by about a factor 1000 for
temperatures around $50\times10^6$~K.  However if the non-enhanced rate for
this reaction is correct, then our models reproduce only part of the Mg-Al
anticorrelation.

One has to conclude that fast rotating massive stars are extremely good
candidates for providing the material from which the long-lived low-mass
stars that we are presently observing formed. Of course the process leading
to the incorporation of this material in new stars is far from being
described here, but at least we have given some reasons to believe that the
interesting ejecta can be retained in the globular cluster. It has also to
be stressed that the stars forming from the wind ejecta of massive stars 
mixed with some pristine interstellar gas would
also present higher helium abundances \citep{SalarisWeiss2006} and of
course high C/N ratios. The values of the \el{C}{12}/\el{C}{13} would be quite
low: all the massive stars have ratio below 5. The corresponding implications 
will be studied in a subsequent work.

\subsection{Why only in GCs?}

To end this section, we can wonder why such a process did only occur in
GCs. The environment of field halo stars is of course very different from
the one of the globular clusters. However, in the past, they probably
share, at least for a short time, a cluster environment. These clusters
were probably much less massive and/or much less concentrated than the
progenitors of the present day globular clusters. This is required for
these clusters to have evaporated, being completely disrupted either by the
energy injections of the first supernovae,\footnote{It is interesting to
  note that the gravitational energy of a globular cluster is of the order
  of $10^{51}$~erg i.e. of the same order of magnitude as the kinetic
  energy emitted by a core collapse supernova explosion.} or by tidal
effects. In that case, star formation triggered by the wind/SN shocks from
evolving massive stars might have been quenched, rendering impossible the
birth of field stars with processed material.

\modif{ Another possibility, also related to the high stellar density in
  GCs, could be that such an environment favors higher rotational
  velocities than less dense environments.  It is however not obvious that
  denser environments would favor high rotational velocities, the stellar
  encounters being able of both spinning up the stars or slowing them down.
  The only thing we can say at the moment is that some observations find
  higher rotational velocities in clusters than in the field:
  \citet{HuangGies2006} determined the projected rotational velocities of
  496 OB stars in clusters within the approximate age range 6-73 Myr.  They
  found that there are fewer slow rotators among the cluster B-type stars
  relative to nearby B stars in the field.  \citet{StromWolff2005} also
  found that stars in $h$ and $\chi$ Persei tend to rotate faster than the
  field stars counterparts. A similar conclusion has been reached by
  \citet{DuftonRyans2006} on the basis of the rotational velocities of
  stars in the two clusters NGC~3293 and NGC~4755. Of course it is
  difficult at the present time to draw firm conclusions about the origin
  of this difference.  Let us just note that these observations may support
  the view according to which stars born in dense environments may have
  faster rotational velocities than stars born in lose aggregates.  }

An alternative to this view is to consider that all the field stars were
born in the progenitors of the present day globular clusters. Let us recall
that field halo stars contain at least hundred times the mass in globular
clusters \citep{Woltjer1975}. The above scenario would therefore imply a
very efficient mechanism that would remove 99\% of stars initially formed
in the progenitor of the present day GC. Moreover, this mechanism should
either be more rapid than the enrichment of the interstellar medium by
massive stars, or should only remove stars with no chemical
inhomogeneities. While it appears difficult to favor such a view, it would
have the advantage of making the stars presenting inhomogeneities a very
small subset of the whole stellar population in a given cluster and thus
alleviate the need of a very flat IMF (PC06).\footnote{Since massive stars
  are preferentially located in the central parts of a young cluster, they
  may have polluted mainly the central regions. The outer part of the GC
  progenitors would at least for a while be composed of non polluted stars
  which may have been stripped off by tidal effects.}

\section{Conclusion}

The two main reasons why massive stars have been discarded in the past in
the context of the GC self-enrichment scenario are related to the fact that
these objects are expected to produce iron and to have fast winds.

In the present paper, we propose the Wind of Fast Rotating Massive Star
scenario in order to explain the chemical inhomogeneities observed in GC
stars. The key point of this scenario is the \textbf{fast rotation} of
\textbf{massive} stars. Each one of these two characteristics is important:
\begin{itemize}
\item {\bf Fast rotation} is needed to remove material from the stellar
  surface and inject it with a low velocity in the interstellar medium. It
  triggers internal mixing which brings to the surface material processed
  in the core. This enables a star to eject material with chemical
  compositions similar to that observed at the surface of GC stars.
\item{\bf Massive stars} have short lifetimes and can release H-synthesized
  material while low-mass stars are still forming in the nascent globular
  cluster. They can, through the wind and SN shocks or through the
  ionization front they produce, trigger star formation in their vicinity,
  being thus able to be at the same time the cause of new star formation
  and the provider of at least part of the material from which the stars
  form. Moreover, it appears that rotating massive stars can lose low-speed
  material only enriched in H-burning products. Low- and intermediate-mass
  stars would not be able to do that, either because they are less
  efficiently mixed by rotation and/or they have more difficulty to reach
  the break-up limit and of course the $\Omega\Gamma$--Limit than the
  massive stars during the main sequence. In addition their central
  temperature is too low to efficiently activate NeNa and MgAl chains
  during the MS. Besides they require a more top-heavy IMF than
  massive stars as explained by PC06.
\end{itemize}

Globular clusters may be suitable environments to form fast rotating
massive stars in the proper range of metallicity. It is interesting to
recall that the WFRMS scenario for explaining the GC abundance anomalies is
only one consequence of fast rotation of massive stars. This kind of stars
might be also interesting for understanding other features as the origin of
the carbon-rich ultra metal-poor stars, the high N/O ratio observed in halo
stars, and the high helium abundance of a part of stars in $\omega$~Cen
\modif{\citep{MaederMeynet2006}}. The anticorrelations observed in GC might
thus be an additional observed consequences of the chemical enrichment
expected from fast rotating massive stars.

\begin{acknowledgements}

\end{acknowledgements}

\bibliographystyle{aa}

\bibliography{BibADS}

\begin{thebibliography}{83}
\expandafter\ifx\csname natexlab\endcsname\relax\def\natexlab#1{#1}\fi

\bibitem[{{Alexander} \& {Ferguson}(1994)}]{AlexanderFerguson1994}
{Alexander}, D.~R. \& {Ferguson}, J.~W. 1994, \apj, 437, 879

\bibitem[{{Angulo} {et~al.}(1999){Angulo}, {Arnould}, {Rayet}, {Descouvemont},
  {Baye}, {Leclercq-Willain}, {Coc}, {Barhoumi}, {Aguer}, {Rolfs}, {Kunz},
  {Hammer}, {Mayer}, {Paradellis}, {Kossionides}, {Chronidou}, {Spyrou},
  {degl'Innocenti}, {Fiorentini}, {Ricci}, {Zavatarelli}, {Providencia},
  {Wolters}, {Soares}, {Grama}, {Rahighi}, {Shotter}, \& {Lamehi
  Rachti}}]{AnguloArnould1999}
{Angulo}, C., {Arnould}, M., {Rayet}, M., {et~al.} 1999, Nuclear Physics A,
  656, 3

\bibitem[{{Arnould} {et~al.}(1999){Arnould}, {Goriely}, \&
  {Jorissen}}]{ArnouldGoriely1999}
{Arnould}, M., {Goriely}, S., \& {Jorissen}, A. 1999, \aap, 347, 572

\bibitem[{{Brown} \& {Wallerstein}(1993)}]{BrownWallerstein1993}
{Brown}, J.~A. \& {Wallerstein}, G. 1993, \aj, 106, 133

\bibitem[{{Carretta} {et~al.}(2004){Carretta}, {Bragaglia}, \&
  {Cacciari}}]{CarrettaBragaglia2004}
{Carretta}, E., {Bragaglia}, A., \& {Cacciari}, C. 2004, \apjl, 610, L25

\bibitem[{{Carretta} {et~al.}(2003){Carretta}, {Bragaglia}, {Cacciari}, \&
  {Rossetti}}]{CarrettaBragaglia2003}
{Carretta}, E., {Bragaglia}, A., {Cacciari}, C., \& {Rossetti}, E. 2003, \aap,
  410, 143

\bibitem[{{Carretta} {et~al.}(2005){Carretta}, {Gratton}, {Lucatello},
  {Bragaglia}, \& {Bonifacio}}]{CarrettaGratton2005}
{Carretta}, E., {Gratton}, R.~G., {Lucatello}, S., {Bragaglia}, A., \&
  {Bonifacio}, P. 2005, \aap, 433, 597

\bibitem[{{Chaboyer} \& {Zahn}(1992)}]{ChaboyerZahn1992}
{Chaboyer}, B. \& {Zahn}, J.-P. 1992, \aap, 253, 173

\bibitem[{{Charbonnel}(2005)}]{Charbonnel2005}
{Charbonnel}, C. 2005, in IAU Symposium, ed. V.~{Hill}, P.~{Fran{\c c}ois}, \&
  F.~{Primas}, 347--356

\bibitem[{{Coc} {et~al.}(2004){Coc}, {Vangioni-Flam}, {Descouvemont},
  {Adahchour}, \& {Angulo}}]{CocVangioniFlam2004}
{Coc}, A., {Vangioni-Flam}, E., {Descouvemont}, P., {Adahchour}, A., \&
  {Angulo}, C. 2004, \apj, 600, 544

\bibitem[{{Cohen} {et~al.}(2002){Cohen}, {Briley}, \&
  {Stetson}}]{CohenBriley2002}
{Cohen}, J.~G., {Briley}, M.~M., \& {Stetson}, P.~B. 2002, \aj, 123, 2525

\bibitem[{{Cottrell} \& {Da Costa}(1981)}]{CottrellDaCosta1981}
{Cottrell}, P.~L. \& {Da Costa}, G.~S. 1981, \apjl, 245, L79

\bibitem[{{de Jager} {et~al.}(1988){de Jager}, {Nieuwenhuijzen}, \& {van der
  Hucht}}]{deJagerNieuwenhuijzen1988}
{de Jager}, C., {Nieuwenhuijzen}, H., \& {van der Hucht}, K.~A. 1988, \aaps,
  72, 259

\bibitem[{{Denissenkov} \& {Herwig}(2003)}]{DenissenkovHerwig2003}
{Denissenkov}, P.~A. \& {Herwig}, F. 2003, \apjl, 590, L99

\bibitem[{{Dickens} {et~al.}(1991){Dickens}, {Croke}, {Cannon}, \&
  {Bell}}]{DickensCroke1991}
{Dickens}, R.~J., {Croke}, B.~F.~W., {Cannon}, R.~D., \& {Bell}, R.~A. 1991,
  \nat, 351, 212

\bibitem[{{Dufton} {et~al.}(2006){Dufton}, {Ryans}, {Sim{\'o}n-D{\'{\i}}az},
  {Trundle}, \& {Lennon}}]{DuftonRyans2006}
{Dufton}, P.~L., {Ryans}, R.~S.~I., {Sim{\'o}n-D{\'{\i}}az}, S., {Trundle}, C.,
  \& {Lennon}, D.~J. 2006, \aap, 451, 603

\bibitem[{{Fenner} {et~al.}(2004){Fenner}, {Campbell}, {Karakas}, {Lattanzio},
  \& {Gibson}}]{FennerCampbell2004}
{Fenner}, Y., {Campbell}, S., {Karakas}, A.~I., {Lattanzio}, J.~C., \&
  {Gibson}, B.~K. 2004, \mnras, 353, 789

\bibitem[{{Freeman} \& {Norris}(1981)}]{FreemanNorris1981}
{Freeman}, K.~C. \& {Norris}, J. 1981, \araa, 19, 319

\bibitem[{{Freyer} {et~al.}(2003){Freyer}, {Hensler}, \&
  {Yorke}}]{FreyerHensler2003}
{Freyer}, T., {Hensler}, G., \& {Yorke}, H.~W. 2003, \apj, 594, 888

\bibitem[{{Garcia-Segura} {et~al.}(1996){Garcia-Segura}, {Langer}, \& {Mac
  Low}}]{GarciaSeguraLanger1996}
{Garcia-Segura}, G., {Langer}, N., \& {Mac Low}, M.-M. 1996, \aap, 316, 133

\bibitem[{{Gnedin} {et~al.}(2002){Gnedin}, {Zhao}, {Pringle}, {Fall}, {Livio},
  \& {Meylan}}]{GnedinZhao02}
{Gnedin}, O.~Y., {Zhao}, H., {Pringle}, J.~E., {et~al.} 2002, \apjl, 568, L23

\bibitem[{{Gratton} {et~al.}(2004){Gratton}, {Sneden}, \&
  {Carretta}}]{GrattonSneden2004}
{Gratton}, R., {Sneden}, C., \& {Carretta}, E. 2004, \araa, 42, 385

\bibitem[{{Gratton} {et~al.}(2001){Gratton}, {Bonifacio}, {Bragaglia},
  {Carretta}, {Castellani}, {Centurion}, {Chieffi}, {Claudi}, {Clementini},
  {D'Antona}, {Desidera}, {Fran{\c c}ois}, {Grundahl}, {Lucatello}, {Molaro},
  {Pasquini}, {Sneden}, {Spite}, \& {Straniero}}]{GrattonBonifacio2001}
{Gratton}, R.~G., {Bonifacio}, P., {Bragaglia}, A., {et~al.} 2001, \aap, 369,
  87

\bibitem[{{Gratton} {et~al.}(2000){Gratton}, {Sneden}, {Carretta}, \&
  {Bragaglia}}]{GrattonSneden2000}
{Gratton}, R.~G., {Sneden}, C., {Carretta}, E., \& {Bragaglia}, A. 2000, \aap,
  354, 169

\bibitem[{{Grundahl} {et~al.}(2002){Grundahl}, {Briley}, {Nissen}, \&
  {Feltzing}}]{GrundahlBriley2002}
{Grundahl}, F., {Briley}, M., {Nissen}, P.~E., \& {Feltzing}, S. 2002, \aap,
  385, L14

\bibitem[{{Hale} {et~al.}(2002){Hale}, {Champagne}, {Iliadis}, {Hansper},
  {Powell}, \& {Blackmon}}]{HaleChampagne2002}
{Hale}, S.~E., {Champagne}, A.~E., {Iliadis}, C., {et~al.} 2002, \prc, 65,
  015801

\bibitem[{{Hale} {et~al.}(2004){Hale}, {Champagne}, {Iliadis}, {Hansper},
  {Powell}, \& {Blackmon}}]{HaleChampagne2004}
{Hale}, S.~E., {Champagne}, A.~E., {Iliadis}, C., {et~al.} 2004, \prc, 70,
  045802

\bibitem[{{Harbeck} {et~al.}(2003){Harbeck}, {Smith}, \&
  {Grebel}}]{HarbeckSmith2003}
{Harbeck}, D., {Smith}, G.~H., \& {Grebel}, E.~K. 2003, \aj, 125, 197

\bibitem[{{Heger} \& {Langer}(2000)}]{HegerLanger2000}
{Heger}, A. \& {Langer}, N. 2000, \apj, 544, 1016

\bibitem[{{Herwig}(2004{\natexlab{a}})}]{Herwig2004a}
{Herwig}, F. 2004{\natexlab{a}}, \apj, 605, 425

\bibitem[{{Herwig}(2004{\natexlab{b}})}]{Herwig2004b}
{Herwig}, F. 2004{\natexlab{b}}, \apjs, 155, 651

\bibitem[{{Huang} \& {Gies}(2006)}]{HuangGies2006}
{Huang}, W. \& {Gies}, D.~R. 2006, \apj, 648, 580

\bibitem[{{Iglesias} \& {Rogers}(1996)}]{IglesiasRogers1996}
{Iglesias}, C.~A. \& {Rogers}, F.~J. 1996, \apj, 464, 943

\bibitem[{{Iliadis} {et~al.}(2001){Iliadis}, {D'Auria}, {Starrfield},
  {Thompson}, \& {Wiescher}}]{IliadisStarrfield2001}
{Iliadis}, C., {D'Auria}, J.~M., {Starrfield}, S., {Thompson}, W.~J., \&
  {Wiescher}, M. 2001, \apjs, 134, 151

\bibitem[{{Ivans} {et~al.}(1999){Ivans}, {Sneden}, {Kraft}, {Suntzeff},
  {Smith}, {Langer}, \& {Fulbright}}]{IvansSneden1999}
{Ivans}, I.~I., {Sneden}, C., {Kraft}, R.~P., {et~al.} 1999, \aj, 118, 1273

\bibitem[{{James} {et~al.}(2004{\natexlab{a}}){James}, {Fran{\c c}ois},
  {Bonifacio}, {Bragaglia}, {Carretta}, {Centuri{\'o}n}, {Clementini},
  {Desidera}, {Gratton}, {Grundahl}, {Lucatello}, {Molaro}, {Pasquini},
  {Sneden}, \& {Spite}}]{JamesBonifacio2004a}
{James}, G., {Fran{\c c}ois}, P., {Bonifacio}, P., {et~al.} 2004{\natexlab{a}},
  \aap, 414, 1071

\bibitem[{{James} {et~al.}(2004{\natexlab{b}}){James}, {Fran{\c c}ois},
  {Bonifacio}, {Carretta}, {Gratton}, \& {Spite}}]{JamesBonifacio2004b}
{James}, G., {Fran{\c c}ois}, P., {Bonifacio}, P., {et~al.} 2004{\natexlab{b}},
  \aap, 427, 825

\bibitem[{{Karakas} \& {Lattanzio}(2003)}]{KarakasLattanzio2003}
{Karakas}, A.~I. \& {Lattanzio}, J.~C. 2003, Publications of the Astronomical
  Society of Australia, 20, 279

\bibitem[{{Kraft}(1994)}]{Kraft1994}
{Kraft}, R.~P. 1994, \pasp, 106, 553

\bibitem[{{Kraft} {et~al.}(1992){Kraft}, {Sneden}, {Langer}, \&
  {Prosser}}]{KraftSneden1992}
{Kraft}, R.~P., {Sneden}, C., {Langer}, G.~E., \& {Prosser}, C.~F. 1992, \aj,
  104, 645

\bibitem[{{Kudritzki} \& {Puls}(2000)}]{KudritzkiPuls2000}
{Kudritzki}, R.-P. \& {Puls}, J. 2000, \araa, 38, 613

\bibitem[{{Lamers} {et~al.}(1995){Lamers}, {Snow}, \&
  {Lindholm}}]{LamersSnow1995}
{Lamers}, H.~J.~G.~L.~M., {Snow}, T.~P., \& {Lindholm}, D.~M. 1995, \apj, 455,
  269

\bibitem[{{Langer}(1998)}]{Langer1998}
{Langer}, N. 1998, \aap, 329, 551

\bibitem[{{Maeda} \& {Nomoto}(2003)}]{MaedaNomoto2003}
{Maeda}, K. \& {Nomoto}, K. 2003, \apj, 598, 1163

\bibitem[{{Maeder}(1999)}]{Maeder1999}
{Maeder}, A. 1999, \aap, 347, 185

\bibitem[{{Maeder} \& {Meynet}(2000)}]{MaederMeynet2000}
{Maeder}, A. \& {Meynet}, G. 2000, \aap, 361, 159

\bibitem[{{Maeder} \& {Meynet}(2001)}]{MaederMeynet2001}
{Maeder}, A. \& {Meynet}, G. 2001, \aap, 373, 555

\bibitem[{{Maeder} \& {Meynet}(2006)}]{MaederMeynet2006}
{Maeder}, A. \& {Meynet}, G. 2006, \aap, 448, L37

\bibitem[{{Maeder} \& {Zahn}(1998)}]{MaederZahn1998}
{Maeder}, A. \& {Zahn}, J.-P. 1998, \aap, 334, 1000

\bibitem[{{Meynet} {et~al.}(2006){Meynet}, {Ekstr{\"o}m}, {Maeder}, \&
  {Barblan}}]{MeynetMaeder2006}
{Meynet}, G., {Ekstr{\"o}m}, S., {Maeder}, A., \& {Barblan}, F. 2006, ArXiv
  Astrophysics e-prints

\bibitem[{{Meynet} \& {Maeder}(1997)}]{MeynetMaeder1997}
{Meynet}, G. \& {Maeder}, A. 1997, \aap, 321, 465

\bibitem[{{Meynet} \& {Maeder}(2000)}]{MeynetMaeder2000}
{Meynet}, G. \& {Maeder}, A. 2000, \aap, 361, 101

\bibitem[{{Meynet} \& {Maeder}(2002)}]{MeynetMaeder2002}
{Meynet}, G. \& {Maeder}, A. 2002, \aap, 390, 561

\bibitem[{{Nugis} \& {Lamers}(2000)}]{NugisLamers2000}
{Nugis}, T. \& {Lamers}, H.~J.~G.~L.~M. 2000, \aap, 360, 227

\bibitem[{{Pasquini} {et~al.}(2005){Pasquini}, {Bonifacio}, {Molaro},
  {Francois}, {Spite}, {Gratton}, {Carretta}, \&
  {Wolff}}]{PasquiniBonifacio2005}
{Pasquini}, L., {Bonifacio}, P., {Molaro}, P., {et~al.} 2005, \aap, 441, 549

\bibitem[{{Porter} \& {Rivinius}(2003)}]{PorterRivinius2003}
{Porter}, J.~M. \& {Rivinius}, T. 2003, \pasp, 115, 1153

\bibitem[{{Powell} {et~al.}(1999){Powell}, {Iliadis}, {Champagne}, {Grossmann},
  {Hale}, {Hansper}, \& {McLean}}]{PowellIliadis1999}
{Powell}, D.~C., {Iliadis}, C., {Champagne}, A.~E., {et~al.} 1999, \nphysa,
  660, 349

\bibitem[{{Prantzos} \& {Charbonnel}(2006)}]{PrantzosCharbonnel2006}
{Prantzos}, N. \& {Charbonnel}, C. 2006, \aap, 458, 135

\bibitem[{{Ram{\'{\i}}rez} \& {Cohen}(2002)}]{RamirezCohen2002}
{Ram{\'{\i}}rez}, S.~V. \& {Cohen}, J.~G. 2002, \aj, 123, 3277

\bibitem[{{Ram{\'{\i}}rez} \& {Cohen}(2003)}]{RamirezCohen2003}
{Ram{\'{\i}}rez}, S.~V. \& {Cohen}, J.~G. 2003, \aj, 125, 224

\bibitem[{{Rivinius}(2005)}]{Rivinius2005}
{Rivinius}, T. 2005, in ASP Conf. Ser. 337: The Nature and Evolution of Disks
  Around Hot Stars, ed. R.~{Ignace} \& K.~G. {Gayley}, 178--+

\bibitem[{{Sackmann} \& {Anand}(1970)}]{SackmannAnand1970}
{Sackmann}, I.-J. \& {Anand}, S.~P.~S. 1970, \apj, 162, 105

\bibitem[{{Salaris} {et~al.}(2006){Salaris}, {Weiss}, {Ferguson}, \&
  {Fusilier}}]{SalarisWeiss2006}
{Salaris}, M., {Weiss}, A., {Ferguson}, J.~W., \& {Fusilier}, D.~J. 2006, ArXiv
  Astrophysics e-prints

\bibitem[{{Shetrone}(1996)}]{Shetrone1996II}
{Shetrone}, M.~D. 1996, \aj, 112, 2639

\bibitem[{{Smith}(1987)}]{Smith1987}
{Smith}, G.~H. 1987, \pasp, 99, 67

\bibitem[{{Smith}(2006)}]{Smith2006}
{Smith}, G.~H. 2006, \pasp, in press

\bibitem[{{Sneden}(2005)}]{Sneden2005}
{Sneden}, C. 2005, in IAU Symposium, ed. V.~{Hill}, P.~{Fran{\c c}ois}, \&
  F.~{Primas}, 337--344

\bibitem[{{Sobeck} {et~al.}(2006){Sobeck}, {Ivans}, {Simmerer}, {Sneden},
  {Hoeflich}, {Fulbright}, \& {Kraft}}]{SobeckIvans2006}
{Sobeck}, J.~S., {Ivans}, I.~I., {Simmerer}, J.~A., {et~al.} 2006, \aj, 131,
  2949

\bibitem[{{Strom} {et~al.}(2005){Strom}, {Wolff}, \& {Dror}}]{StromWolff2005}
{Strom}, S.~E., {Wolff}, S.~C., \& {Dror}, D.~H.~A. 2005, \aj, 129, 809

\bibitem[{{Talon}(2004)}]{Talon2004}
{Talon}, S. 2004, in IAU Symposium, ed. A.~{Maeder} \& P.~{Eenens}, 336--+

\bibitem[{{Townsend} {et~al.}(2004){Townsend}, {Owocki}, \&
  {Howarth}}]{TownsendOwocki2004}
{Townsend}, R.~H.~D., {Owocki}, S.~P., \& {Howarth}, I.~D. 2004, \mnras, 350,
  189

\bibitem[{{Ventura} \& {D'Antona}(2005{\natexlab{a}})}]{VenturaDAntona2005a}
{Ventura}, P. \& {D'Antona}, F. 2005{\natexlab{a}}, \aap, 431, 279

\bibitem[{{Ventura} \& {D'Antona}(2005{\natexlab{b}})}]{VenturaDAntona2005b}
{Ventura}, P. \& {D'Antona}, F. 2005{\natexlab{b}}, \aap, 439, 1075

\bibitem[{{Ventura} \& {D'Antona}(2005{\natexlab{c}})}]{VenturaDAntona2005c}
{Ventura}, P. \& {D'Antona}, F. 2005{\natexlab{c}}, \apjl, 635, L149

\bibitem[{{Ventura} \& {D'Antona}(2006)}]{VenturaDAntona2006}
{Ventura}, P. \& {D'Antona}, F. 2006, \aap, 457, 995

\bibitem[{{Ventura} {et~al.}(2002){Ventura}, {D'Antona}, \&
  {Mazzitelli}}]{VenturaDAntona2002}
{Ventura}, P., {D'Antona}, F., \& {Mazzitelli}, I. 2002, \aap, 393, 215

\bibitem[{{Ventura} {et~al.}(2001){Ventura}, {D'Antona}, {Mazzitelli}, \&
  {Gratton}}]{VenturaDAntona2001}
{Ventura}, P., {D'Antona}, F., {Mazzitelli}, I., \& {Gratton}, R. 2001, \apjl,
  550, L65

\bibitem[{{Wallerstein} {et~al.}(1987){Wallerstein}, {Myckky-Leep}, \&
  {Oke}}]{WallersteinMyckkyLeep1987}
{Wallerstein}, G., {Myckky-Leep}, E., \& {Oke}, J.~B. 1987, \aj, 94, 523

\bibitem[{{Woltjer}(1975)}]{Woltjer1975}
{Woltjer}, L. 1975, \aap, 42, 109

\bibitem[{{Yong} {et~al.}(2006){Yong}, {Aoki}, \& {Lambert}}]{YongAoki2006}
{Yong}, D., {Aoki}, W., \& {Lambert}, D.~L. 2006, \apj, 638, 1018

\bibitem[{{Yong} {et~al.}(2003){Yong}, {Grundahl}, {Lambert}, {Nissen}, \&
  {Shetrone}}]{YongGrundahl2003}
{Yong}, D., {Grundahl}, F., {Lambert}, D.~L., {Nissen}, P.~E., \& {Shetrone},
  M.~D. 2003, \aap, 402, 985

\bibitem[{{Yong} {et~al.}(2005){Yong}, {Grundahl}, {Nissen}, {Jensen}, \&
  {Lambert}}]{YongGrundahl2005}
{Yong}, D., {Grundahl}, F., {Nissen}, P.~E., {Jensen}, H.~R., \& {Lambert},
  D.~L. 2005, \aap, 438, 875

\bibitem[{{Zahn}(1992)}]{Zahn1992}
{Zahn}, J.-P. 1992, \aap, 265, 115

\end{thebibliography}

\end{document}